\documentclass[11pt,3p,times]{elsarticle}

\usepackage{amsmath,amsfonts,amssymb}
\usepackage{amsthm}
\usepackage{bm}
\usepackage{graphicx,epstopdf}
\usepackage[position=top,labelfont=normalfont,textfont=normalfont,singlelinecheck=off,justification=raggedright]{subfig}
\usepackage{floatrow}
\usepackage[dvipsnames]{xcolor}
\usepackage{setspace}
\usepackage{enumerate,etaremune}
\usepackage{booktabs}
\usepackage{tabularx,multirow}
\usepackage{framed}
\usepackage{hhline}
\usepackage{url}
\usepackage[unicode,bookmarks=false]{hyperref}
\hypersetup{
  colorlinks,
  citecolor=Blue,linkcolor=Blue,urlcolor=Blue
}

\usepackage{lineno}

\usepackage[textsize=footnotesize,linecolor=red,backgroundcolor=red!25,bordercolor=red]
  {todonotes}

\usepackage{tikz}
\usetikzlibrary{shapes.geometric}
\usetikzlibrary{shapes,arrows}
\usetikzlibrary{plotmarks}

\usetikzlibrary{calc}

\usepackage{physics}

\usepackage{algorithm}
\usepackage{algorithmicx,algpseudocode}
\usepackage{cases}
\newcommand{\eg}{{\it e.g.}}
\newcommand{\ie}{{\it i.e.}}
\newcommand{\cf}{{\it cf.}}
\newcommand{\etal}{{\it et al.}}
\newcommand{\tensor}[1]{\bm{#1}}
\newcommand{\stress}{\sigma}
\newcommand{\strain}{\varepsilon}
\newcommand{\tstress}{\tensor{\stress}}
\newcommand{\tstrain}{\tensor{\strain}}

\newcommand{\jump}[1]{\lbrack\!\lbrack #1 \rbrack\!\rbrack}

\newcommand{\pd}{\partial}

\newcommand{\el}{\mathrm{e}}
\newcommand{\pl}{\mathrm{p}}

\newcommand{\rn}[1]{\uppercase\expandafter{\romannumeral #1\relax}}
\newcommand{\cn}{\mathrm{N}}

\newcommand{\bulk}{\mathrm{bulk}}
\newcommand{\inter}{\mathrm{interface}}

\let\grad\relax
\DeclareMathOperator{\grad}{\nabla}
\let\diver\relax
\DeclareMathOperator{\diver}{\nabla\cdot}
\DeclareMathOperator{\symgrad}{\nabla^{s}}

\DeclareMathOperator{\dyadic}{\otimes}
\newsavebox{\dotbox}

\theoremstyle{remark}
\newtheorem{remark}{Remark}

\newcommand{\revised}[1]{{\color{black} #1}}
\newcolumntype{L}[1]{>{\raggedright\let\newline\\arraybackslash\hspace{0pt}}m{#1}}
\newcolumntype{C}[1]{>{\centering\let\newline\\arraybackslash\hspace{0pt}}m{#1}}
\newcolumntype{R}[1]{>{\raggedleft\let\newline\\arraybackslash\hspace{0pt}}m{#1}}

\allowdisplaybreaks

\biboptions{sort&compress,square,comma,numbers}
\AtBeginDocument{\hypersetup{citecolor=MidnightBlue,linkcolor=MidnightBlue,urlcolor=MidnightBlue}}

\usepackage{etoolbox}
\makeatletter
\patchcmd{\ps@pprintTitle}% <cmd>
{Preprint submitted to}% <search>
{~}% <replace>
{}{}% <succes><failure>
\makeatother
\begin{document}

\begin{frontmatter}

\title{Phase-field modeling of rock fractures with roughness}

\author[HKU]{Fan Fei}
\author[HKU,KAIST]{Jinhyun Choo\corref{corr}}
\ead{jchoo@hku.hk}
\author[HKU]{Chong Liu}
\author[LLNL]{Joshua A. White}

\cortext[corr]{Corresponding Author}

\address[HKU]{Department of Civil Engineering, The University of Hong Kong, Hong Kong}
\address[KAIST]{Department of Civil and Environmental Engineering, KAIST, South Korea}
\address[LLNL]{Atmospheric, Earth, and Energy Division, Lawrence Livermore National Laboratory, United States}

\journal{~}

\begin{abstract}
Phase-field modeling---a continuous approach to discontinuities---is gaining popularity for simulating rock fractures due to its ability to handle complex, discontinuous geometry without an explicit surface tracking algorithm.
None of the existing phase-field models, however, incorporates the impact of surface roughness on the mechanical response of fractures---such as elastic deformability and shear-induced dilation---despite the importance of this behavior for subsurface systems.
To fill this gap, here we introduce the first framework for phase-field modeling of rough rock fractures.
The framework transforms a displacement-jump-based discrete constitutive model for discontinuities into a strain-based continuous model, without any additional parameter, and then casts it into a phase-field formulation for frictional interfaces.
We illustrate the framework by constructing a particular phase-field form employing a rock joint model originally formulated for discrete modeling.
The results obtained by the new formulation show excellent agreement with those of a well-established discrete method for a variety of problems ranging from shearing of a single discontinuity to compression of fractured rocks.
It is further demonstrated that the phase-field formulation can well simulate complex crack growth from rough discontinuities.
Consequently, our phase-field framework provides an unprecedented bridge between a discrete constitutive model for rough discontinuities---common in rock mechanics---and the continuous finite element method---standard in computational mechanics---without any algorithm to explicitly represent discontinuity geometry.
\end{abstract}


\end{frontmatter}

% \linenumbers

% SECTION 1
% ------------------------------------------------------------------------------
\section{Introduction}
\label{sec:intro}
Rock fractures are pervasive in natural and engineered subsurface systems.
The mechanical behavior of rock fractures not only controls the performance of many geotechnical structures such as slopes and tunnels (\eg~\cite{barton2000tbm,el2009deep,borja2016rock}), but it plays an important role in the operation of subsurface energy technologies such as hydraulic stimulation, nuclear waste disposal, enhanced geothermal systems, and geologic carbon storage (\eg~\cite{white2014geomechanical,barton2020review,lepillier2020variational,fu2021close,fu2021thermo}).

Traditionally, numerical models have treated rock fractures as discrete lower-dimensional entities: lines in two-dimensional domains, and surfaces in three-dimensional domains.
While such discrete modeling of rock fractures has been standard in geomechanical research and practice, it inevitably requires one to explicitly represent the discontinuous geometry of fractures.
Unfortunately, this is a challenging endeavor whenever the geometry of discontinuities is nontrivial, not to mention evolving.

In recent years, phase-field modeling has emerged as a robust method to handle complex fracture geometry without any explicit representation.
This method approximates discontinuous geometry \emph{diffusely} using a continuous field variable called the phase field.
After being developed for general solids~\cite{bourdin2008variational,miehe2010thermodynamically,borden2012phase}, phase-field modeling of fracture has been adopted to simulate rock fracture in a variety of contexts, from hydraulic fracturing to cracking from preexisting flaws (\eg~\cite{lee2016pressure,zhang2017modification,bryant2018mixed,choo2018coupled,ha2018liquid,fei2021double}).

Nevertheless, the vast majority of phase-field simulations of rock fracture have completely disregarded physical phenomena emanating from the \emph{roughness} of fracture.
The surfaces of rock fractures are often rough due to asperities at multiple scales, interlocking them under in-situ stress conditions.
The degree of roughness is so important to subsurface system behavior that it has motivated the development and widespread use of the joint roughness coefficient (JRC)~\cite{barton1977shear,barton1982modelling,barton1985strength} and similar measures in rock mechanics.
Asperity interlocking and relative slip leads to important characteristics that may be absent in fractures in other materials.
First, rock fractures permit a finite amount of elastic deformation under both normal and shear loading due to asperity compressibility.
Second, surface roughness can give rise to a significant amount of dilation when the fracture is sheared.
This dilation in turn contributes to the shear strength of the fracture, making the peak shear strength substantially greater than the residual strength.
It is also noted that the dilation and friction in a rough fracture may depend strongly on the shearing rate and state of the fracture.
Third, the roughness of fracture can evolve by cyclic loading, asperity damage, and other multiphysical phenomena, which affects all of the aforementioned characteristics.

Only very recently, frictional effects have been incorporated into phase-field modeling of fracture.
Fei and Choo~\cite{fei2020phasea} were the first to develop a phase-field formulation for cracks and interfaces with frictional contact, whereby the responses of a diffusely-approximated discontinuity under different contact conditions (open, stick, and slip) are modeled based on stress decomposition in an interface-oriented coordinate system.
Building on this work, the authors then proposed phase-field formulations for frictional shear fracture~\cite{fei2020phaseb} and mixed-mode fracture in quasi-brittle materials~\cite{fei2021double}, in which the contribution of frictional energy to fracturing is considered in a manner consistent with the fracture mechanics theory of Palmer and Rice~\cite{palmer1973growth}.
Meanwhile, Bryant and Sun~\cite{bryant2021phase} have extended the work of Fei and Choo~\cite{fei2020phasea} to incorporate variable friction under non-isothermal conditions.

None of the existing phase-field models, however, can accommodate other aspects of roughness effects.
Among them, elastic deformability and shear-induced dilation are deemed critical as they have significant impacts.
For example, shear-induced dilation in rough fractures can significantly increases the fracture permeability~\cite{barton1985strength,olsson2001improved}.
Hydro-shearing, which injects fluid to exploit this mechanism of permeability enhancement, is receiving growing attention in shale gas production~\cite{hakso2019relation} as well as playing a central role in paving the way to enhanced geothermal systems~\cite{petty2013improving,gischig2015hydro,rinaldi2019joint}.

In discrete methods for discontinuities, the mechanical behavior of a rough rock fracture is modeled by a constitutive law that relates the displacement jump (relative displacement) across the fracture surfaces---the kinematic quantity measured in a shear-box test---to the surface traction.
A large number of such constitutive models have been proposed to capture an array of roughness-induced phenomena such as elasticity, variable friction, shear-induced dilation, and asperity degradation, in a phenomenological and/or a physically-motivated manner (\eg~\cite{barton1977shear,heuze1981new,gens1990constitutive,saeb1990modelling,plesha1987constitutive,white2014anisotropic}).

However, these constitutive models of rough rock fractures---formulated based on the displacement jump across discrete surfaces---are incompatible with diffusely-approximated fractures in phase-field modeling.
Overcoming this incompatibility requires one to transform a displacement-based constitutive model into a strain-based model compatible with the phase-field approximation.

In this work, we introduce the first framework for phase-field modeling of rough fractures.
The framework transforms a displacement-jump-based discrete constitutive model for discontinuities into a strain-based continuous model, and then cast it into a phase-field formulation for frictional interfaces.
Notably, no additional parameter is introduced to the existing phase-field formulation.

The paper is organized as follows.
In Section~\ref{sec:phase-field-formulation}, we adopt a general phase-field formulation for fractured solids subject to compressive stress, in which frictional contact is handled by the stress decomposition scheme proposed by Fei and Choo~\cite{fei2020phasea}.
In Section~\ref{sec:constitutive}, we formulate a strain-based kinematic description of rough rock fractures in the phase-field setting and then develop a general methodology to transform a displacement-based discontinuity model into a strain-based model.
For the purpose of illustration, we construct a particular version of the framework employing an existing displacement-based model for rough rock joints~\cite{white2014anisotropic}.
In Section~\ref{sec:discretization}, we describe how to solve the proposed formulation using a standard nonlinear finite element method in conjunction with Newton's method.
In Section~\ref{sec:examples}, we first verify the proposed phase-field formulation by comparing its results with those obtained by a well-established discrete method (the extended finite element method), using various problems involving a single fracture, non-intersecting fractures, and intersecting fractures.
We then apply the phase-field formulation to simulate the nucleation and propagation of fractures from rough discontinuities.
We conclude the work in Section~\ref{sec:closure}.

% SECTION 2
% ------------------------------------------------------------------------------
\section{Phase-field formulation}
\label{sec:phase-field-formulation}
This section recapitulates a general phase-field formulation for fractured solids possibly subject to compressive stress.
We first describe a standard phase-field approximation of discontinuities and associated governing equations.
We then explain the stress decomposition scheme proposed by Fei and Choo~\cite{fei2020phasea}, which is a well-verified way to incorporate frictional contact in the phase-field setting.
Without loss of generality, we assume infinitesimal deformation and quasi-static conditions.

\subsection{Phase-field approximation and governing equations}
Consider the solid body $\Omega \in \mathbb{R}^\mathrm{dim}$ ($\mathrm{dim}=2,3$) with boundary $\pd \Omega$.
The boundary is suitably decomposed into the displacement (Dirichlet) boundary $\pd_{u}\Omega$ and the traction (Neumann) boundary $\pd_{t}\Omega$ such that $\pd_{u}\Omega \cap \pd_{t}\Omega = \emptyset$ and $\overline{\pd_{u}\Omega \cup \pd_{t}\Omega} = \pd\Omega$.
The body may contain a set of discontinuities denoted by $\Gamma$, which may evolve in the time domain $\mathcal{T}$.

\begin{figure}[h!]
  \centering
  \includegraphics[width=0.9\textwidth]{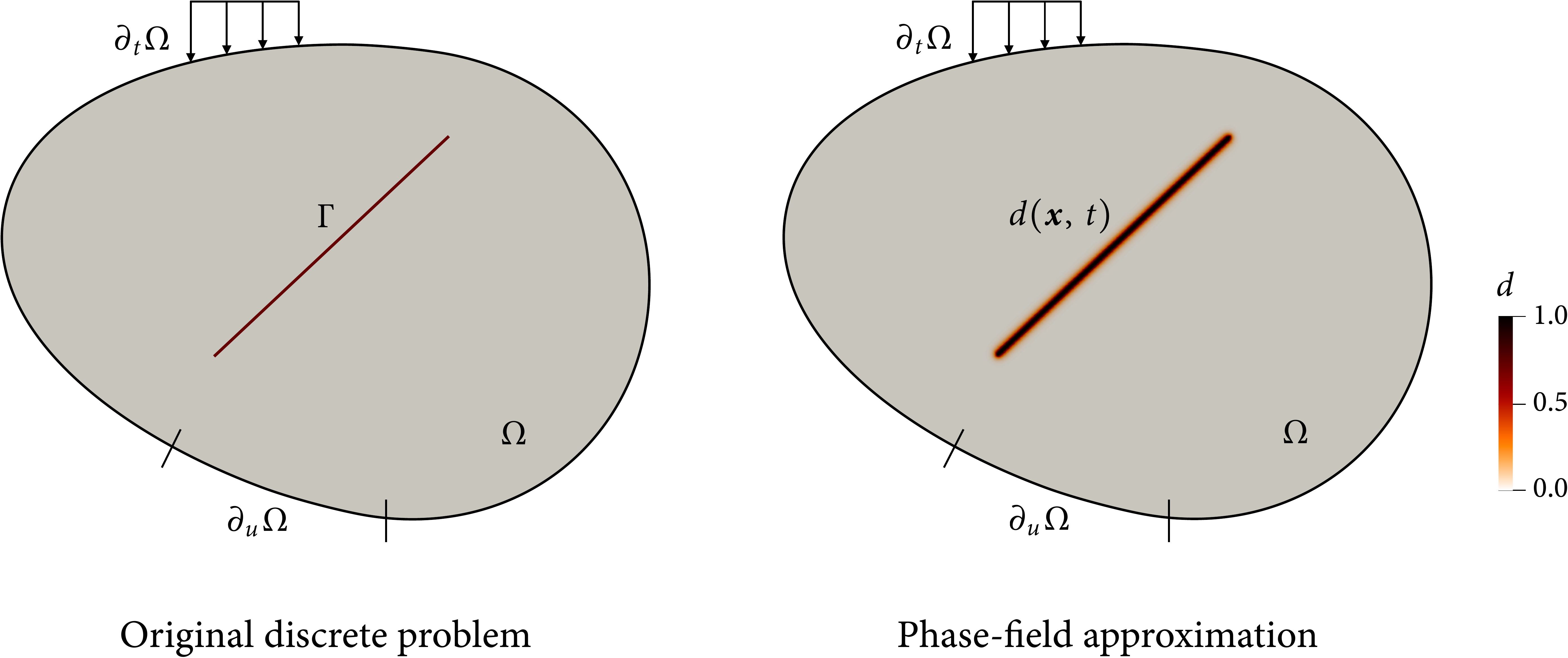}
  \caption{Phase-field approximation of fracture. The discontinuity $\Gamma$ on the left hand side is diffusively represented by the phase-field variable $d$ on the right hand side.}
  \label{fig:phase-field-approximation}
\end{figure}

Figure~\ref{fig:phase-field-approximation} illustrates how phase-field modeling diffusely approximates the original discrete problem.
The approximation begins by introducing the phase-field variable, $d \in [0,\,1]$, where $d = 0$ indicates the undamaged (bulk) region and $d=1$ indicates the fully cracked (interface) region.
We then introduce a crack density functional, $\Gamma_{d}(d, \grad d)$, that satisfies
\begin{equation}
	\Gamma \approx \int_{\Omega} \Gamma_{d}(d, \grad d) \: \dd V .
\end{equation}
A general form of $\Gamma_{d}(d, \grad d)$ is given by
\begin{equation}
  \Gamma_{d}(d,\grad d) := \dfrac{1}{c_0}\left[\dfrac{\alpha(d)}{L} + L\grad{d}\cdot\grad{d}\right], \quad \text{where}\;\;
  c_{0} := 4\int_{0}^{1}\sqrt{\alpha(s)}\,\dd s.
  \label{eq:crack-density-functional-general}
\end{equation}
Here, $L$ is the phase-field length parameter controlling the width of diffuse approximation.
The specific form of $\Gamma_{d}(d,\grad d)$ is determined by the choice of $\alpha(d)$.
Among a few forms of $\alpha(d)$ proposed in the literature~\cite{wu2017unified,geelen2019phase}, here we choose $\alpha(d)=d$ for its relative simplicity and its relation to cohesive fracture~\cite{geelen2019phase}.
We note that other choices of $\alpha(d)$ would equally be valid for the purpose of diffuse approximation of stationary cracks (see Fei and Choo~\cite{fei2020phasea} where $\alpha(d)=d^2$ is used), but for simulating evolving fractures, they would lead to different results.
Substituting $\alpha(d)=d$ into Eq.~\eqref{eq:crack-density-functional-general} gives
\begin{equation}
	\Gamma_{d}(d,\grad d) = \dfrac{3}{8}\qty[\dfrac{d}{L} + L\grad{d}\cdot\grad{d}]. \label{eq:crack-density-function-lorentz}
\end{equation}

The phase-field formulation gives rise to two governing equations---the linear momentum balance equation and the phase-field evolution equation---of which the primary variables are the displacement vector, $\tensor{u}$, and the phase field, $d$.
For brevity, we omit their derivations and refer to the relevant literature (\eg~\cite{geelen2019phase,fei2020phaseb}).
The governing equations may be written as
\begin{linenomath}
\begin{align}
    \diver \tstress(\tensor{u},d) + \rho\tensor{g} = \tensor{0} &\quad  \text{in}\:\: \Omega \times \mathcal{T} ,
    \label{eq:momentum-balance}\\
    -g'(d) \mathcal{H}^{+}(\tensor{u}) + \dfrac{3\mathcal{G}_{c}}{8L} \left(1 - L^2 \diver \grad d\right) = 0 &\quad \text{in}\:\: \Omega\times \mathcal{T}.
    \label{eq:phase-field-evolution}
\end{align}
\end{linenomath}
In the momentum balance equation~\eqref{eq:momentum-balance}, $\tstress$ is the Cauchy stress tensor, $\rho$ is the mass density, and $\tensor{g}$ is the gravitational acceleration vector.
In the phase-field evolution equation~\eqref{eq:phase-field-evolution}, $\mathcal{G}_{c}$ is the critical fracture energy, $\mathcal{H}^{+}$ is the crack driving force, and  $g(d)$ is the degradation function.
It is noted that the specific forms of $\mathcal{H}^{+}$ and $g(d)$ should be chosen in accordance with the selected form of the crack density functional, $\Gamma_d(d,\grad d)$.
For the particular form of $\Gamma_d(d,\grad d)$ in Eq.~\eqref{eq:crack-density-function-lorentz}, $g(d)$ is given by\revised{~\cite{lorentz2011convergence,lorentz2011gradient,geelen2019phase}}
\begin{equation}
  g(d) = \dfrac{(1 - d)^{2}}{(1 - d)^{2} + m d(1 - d)}, \quad \text{with} \:\: m = \dfrac{3\mathcal{G}_{c}}{4L}\dfrac{1}{\mathcal{H}_{t}}.
\end{equation}
Here, $\mathcal{H}_{t}$ is defined as the threshold value of the crack driving force at the peak material strength.
\revised{This threshold value is assigned as the initial value of $\mathcal{H}^{+}$ at intact ($d=0$) material points, ensuring that $d$ is equal to or greater than zero. (Without such a constraint, $d$ does not have a lower bound when Eq.~\eqref{eq:crack-density-function-lorentz} is used as the crack density functional, see Gerasimov and De Lorenzis~\cite{gerasimov2019penalization}.)}
The specific expressions for $\mathcal{H}^{+}$ and $\mathcal{H}_{t}$ depend also on how the stress tensor is decomposed into its crack driving and non-driving parts. So they will be discussed after presenting the stress decomposition scheme.

To complete the problem statement, we write the boundary conditions as
\begin{linenomath}
\begin{align}
		\tensor{u} = \hat{\tensor{u}} &\quad \text{on}\:\: \pd_{u} \Omega \times \mathcal{T} ,  \\
		\tstress \cdot \tensor{v} = \hat{\tensor{t}} &\quad  \text{on}\:\: \pd_{t} \Omega \times \mathcal{T} ,  \\
		\grad d \cdot \tensor{v} = 0 &\quad \text{on}\:\: \pd \Omega \times \mathcal{T} ,
\end{align}
\end{linenomath}
where $\hat{\tensor{u}}$ and $\hat{\tensor{t}}$ are the prescribed displacement and traction vectors on the boundary, respectively, and $\tensor{v}$ is the unit vector outward normal to the boundary.
The initial conditions are given by
\begin{linenomath}
\begin{align}
		\tensor{u} \rvert_{t = 0} = \tensor{u}_0 &\quad \text{in} \:\: \Omega , \\
		d \rvert_{t = 0} = d_0 &\quad \text{in} \:\: \Omega ,
\end{align}
\end{linenomath}
where $\tensor{u}_{0}$ and $d_{0}$ are the initial displacement and phase-field solutions, respectively.
\smallskip

\begin{remark}
When dealing with stationary fractures---a fairly common scenario in rock mechanics applications---Eq.~\eqref{eq:phase-field-evolution} needs to be solved only once for obtaining an initial phase field that suitably represents the preexisting fractures. 
It is noted that such stationary fractures have also been well modeled by embedded discontinuity methods (\eg~\cite{belytschko2001arbitrary,rivas2019two,cusini2021simulation}).
Compared with these methods, the phase-field method entails more computation time as it requires a quite fine discretization around the discontinuity. 
However, the phase-field method lends itself to a much easier implementation because it uses the standard basis (shape) functions whereas embedded discontinuity methods commonly require enrichment of basis functions.
\end{remark}

\subsection{Stress decomposition incorporating frictional contact}
Given that our interest is in fracture under compressive loads, we adopt the stress decomposition scheme proposed by Fei and Choo~\cite{fei2020phasea}, which is a verified way to incorporate frictional contact into phase-field-approximate discontinuities.
It calculates the stress tensor $\tstress$ as the weighted average of the stress in the rock matrix (bulk region), $\tstress_{m}$, and the stress in the fracture (interface region), $\tstress_{f}$,\footnote{$\tstress_{m}$ and $\tstress_{f}$ correspond to $\tstress_{\bulk}$ and $\tstress_{\inter}$ in Fei and Choo~\cite{fei2020phasea}, respectively.} with the weighting done according to the degradation function, $g(d)$.
Specifically,
\begin{equation}
  \tstress = g(d)\tstress_{m} + [1 - g(d)]\tstress_{f}.
  \label{eq:stress-partition}
\end{equation}

The matrix stress tensor can be calculated using a standard continuum constitutive relationship.
Here we assume that the rock matrix is linear elastic, such that the constitutive relationship is given by
\begin{equation}
	\tstress_{m} = \mathbb{C}^{\el}_{m}:\tstrain,
\end{equation}
where $\tstrain$ is the infinitesimal strain tensor, and $\mathbb{C}^{\el}$ is the elastic stiffness tensor.
The linear elastic stiffness tensor can be expressed using the bulk modulus, $K_m$, and the shear modulus, $G_m$ of the rock matrix as
\begin{equation}
	\mathbb{C}^{\el}_{m} = K_{m} \tensor{1} \dyadic \tensor{1} + 2G_{m} \left(\mathbb{I} - \dfrac{1}{3}\tensor{1} \dyadic \tensor{1} \right),
\end{equation}
with $\mathbb{I}$ and $\tensor{1}$ denoting the fourth-order symmetric identity tensor and the second-order identity tensor, respectively.

\begin{figure}[h!]
  \centering
  \includegraphics[width=0.9\textwidth]{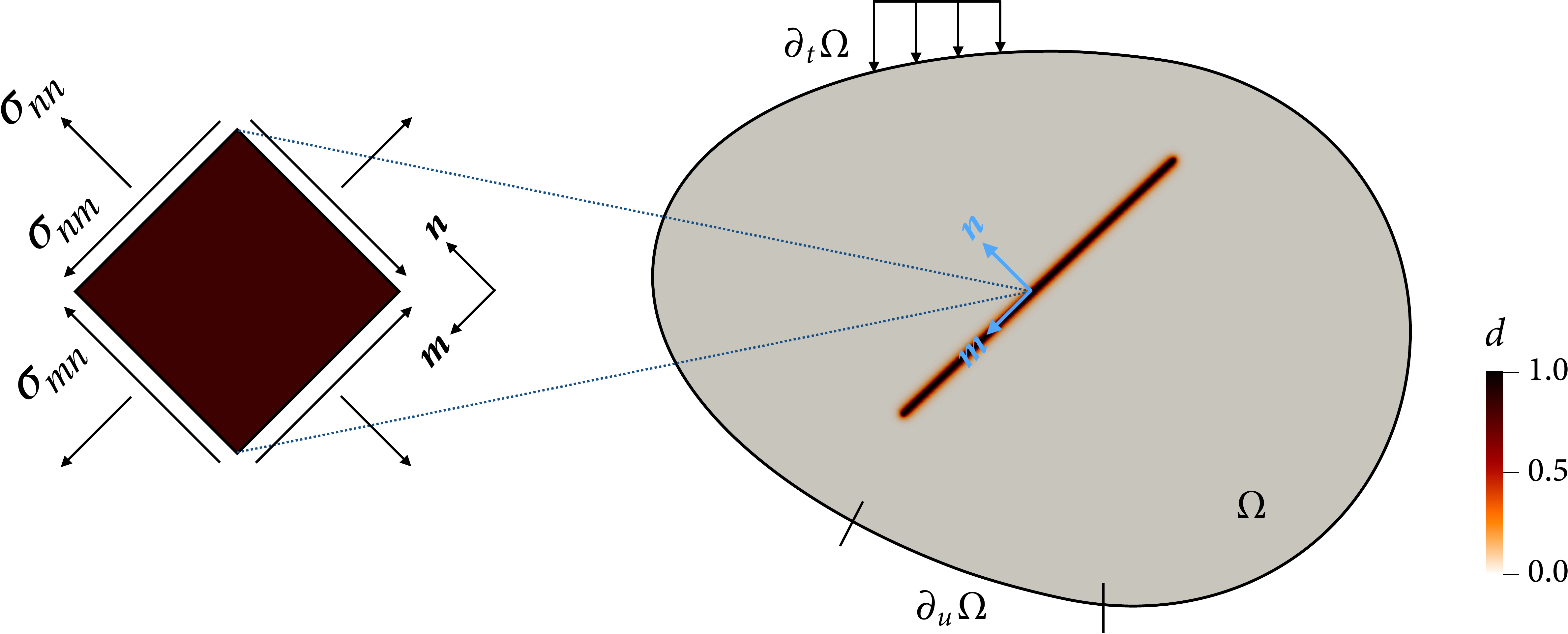}
  \caption{The interface-oriented coordinate system in the phase-field model. Unit vectors $\tensor{n}$ and $\tensor{m}$ denote the normal and tangent to the interface, respectively.}
  \label{fig:interface-coord}
\end{figure}

The fracture stress tensor $\tstress_{f}$, which is nonzero whenever $d>0$, is computed depending on the contact condition at the material point.
To determine the contact condition, we introduce an interface-oriented coordinate system.  
In 2D, for example, the coordinate system is defined with the unit normal vector $\tensor{n}$ and the unit tangent vector $\tensor{m}$ of the discontinuity, see Fig.~\ref{fig:interface-coord}. 
In this interface-oriented coordinate system, the normal strain $\strain_{\cn}$ is calculated as
\begin{equation}
  \strain_{\cn} := \tstrain:(\tensor{n} \dyadic \tensor{n}) .
  \label{eq:normal-eps}
\end{equation}
We can distinguish between open and closed contact conditions based on the sign of $\strain_{\cn}$: open if $\strain_{\cn} > 0$, and closed otherwise.
When the crack is open, $\tstress_{f} = \tensor{0}$.
On the other hand, when the crack is closed, $\tstress_{f}$ should be calculated such that it incorporates the contact behavior of the discontinuity.

Having decomposed the stress tensor as above, let us now determine the specific expression for the crack driving force, $\mathcal{H}^{+}$.
Considering that rocks are quasi-brittle (rather than perfectly brittle) and cracks from rough discontinuities are mostly tensile~\cite{asadizadeh2019mechanical}, here we adopt the crack driving force for cohesive tensile fracture derived from a directionally decomposed stress tensor~\cite{fei2021double}.
Employing the popular approach to crack irreversibility that uses the maximum $\mathcal{H}^{+}$ in history~\cite{miehe2010phase}, the crack driving force is given by 
\begin{linenomath}
\begin{align}
	\mathcal{H}^{+} = \max \left\{ \mathcal{H}_{t} , \, \dfrac{1}{2M}\left[ \max_{t \in [0,t]} \stress_{1,m}(t) \right]^2 \right\}, \quad \text{with} \:\: \mathcal{H}_{t} = \dfrac{1}{2M} \stress^{2}_{p}.
  \label{eq:crack-driving-force}
\end{align}
\end{linenomath}
Here, $M_{m} := K_{m} + (4/3)G_{m}$ is the constrained modulus of the matrix, $\stress_{1,m}$ is the maximum (tensile) principal stress in the matrix, and $\stress_{p}$ is the peak tensile strength.
The detailed derivation of Eq.~\eqref{eq:crack-driving-force} can be found from Fei and Choo~\cite{fei2021double}.
Note that one may also arrive at the same expression for the crack driving force by extending the crack driving force in Steinke and Kaliske~\cite{steinke2019phase}, where the same type of directional stress decomposition is used for brittle tensile fracture, to cohesive tensile fracture as done in Geelen \etal~\cite{geelen2019phase}.

In previous works where the above stress decomposition scheme is employed~\cite{fei2020phasea,fei2020phaseb,fei2021double,bryant2021phase}, the contact behavior is modeled using the standard Coulomb friction law.
However, as explained in the Introduction, the Coulomb friction law does not incorporate a complete array of important features that emanate from the rough nature of rock fractures.
In the following section, we develop a general constitutive framework that allows one to cast a standard model for rough rock fractures---originally formulated for discrete modeling of discontinuities---into $\tstress_{f}$ in diffusive phase-field modeling.

% SECTION 3
% ------------------------------------------------------------------------------
\section{Constitutive framework for rough fractures in phase-field modeling}
\label{sec:constitutive}

In this section, we develop a framework for transforming a displacement-jump-based constitutive model for discontinuities into a strain-based model.
While the framework is general, we construct a particular version of the framework for illustration purposes, employing a specific constitutive model for rough rock joints~\cite{white2014anisotropic}.
To avoid ambiguity, the discussion in this section pertains to an interface material point where $0<d\leq 1$.

\subsection{Kinematics of rough fracture}
\label{sec:kinematics}

In discrete approaches to modeling discontinuities inside continuous materials, the displacement field is often decomposed into its continous part, $\tensor{u}_{c}$, and its discontinuous part---the displacement jump---$\jump{\bm{u}} := \tensor{u}^+-\tensor{u}^-$.
Mathematically, the decomposition can be written as
\begin{equation}
	\tensor{u} = \tensor{u}_{c} + H_{\Gamma}(\tensor{x}) \jump{\bm{u}} ,
  \label{eq:disp-decompose}
\end{equation}
where $H(\tensor{x})$ is the Heaviside function, defined as 
\begin{equation}
	H_{\Gamma}(\tensor{x}) = 
	\left \{
	\begin{array}{ll}
		1 & \text{if} \:\: \tensor{x} \in \Omega_{+}, \\ 
		0 & \text{if} \:\: \tensor{x} \in \Omega_{-},
	\end{array}
	\right . 
\end{equation}
with $\Omega_{+}$ and $\Omega_{-}$ denoting the subdomains that are separated by the discontinuity $\Gamma$. 

To model the mechanical behavior of discontinuities, a constitutive law must be introduced relating the displacement jump $\jump{\bm{u}}$ to the mobilized traction $\tensor{t}$ on the surface.  
This constitutive law is often expressed in a generic rate form as
\begin{equation}
  \dot{\tensor{t}} = \mathbb{D} \cdot \dot{\jump{\bm{u}}} \,,
\end{equation}
where $\mathbb{D}$ is the instantaneous tangent stiffness of the fracture.  It is convenient to develop such models using the surface aligned coordinate system in Fig.~\ref{fig:interface-coord}. We therefore decompose the displacement jump vector into its normal and tangential components as
\begin{equation}
  \jump{\bm{u}} = u_{f} \tensor{n} + v_{f} \tensor{m} ,  \label{eq:disp-decompose-direction}
\end{equation}
where scalars $u_{f}$ and $v_{f}$ denote the magnitude of the normal closure and tangential slip, respectively.

Our first task is to devise a way to model the kinematics of fracture modeled by the phase-field method, which lacks an explicit representation of the displacement jump $\jump{\bm{u}}$.  
To begin, we calculate the strain tensor as the symmetric gradient of the decomposed displacement field~\eqref{eq:disp-decompose}
\begin{equation}
	\tstrain := \symgrad \tensor{u} = \tstrain_{c} + \tstrain_{f},  
  \label{eq:strain-decomposition-total}
\end{equation}
where
\begin{align}
  \tstrain_{c} &:= \symgrad \tensor{u}_{c}, \\
  \tstrain_{f} &:= H_{\Gamma}(\tensor{x})\symgrad \jump{\bm{u}} + \dfrac{1}{2}(\jump{\bm{u}} \dyadic \tensor{n} + \tensor{n} \dyadic \jump{\bm{u}}) \delta_{\Gamma} (\tensor{x}) ,  
  \label{eq:strain-f}
\end{align}
are the continuous and discontinuous parts of the strain tensor, respectively, and $\delta_{\Gamma}(\tensor{x})$ is the Dirac delta function emanating from the gradient of the Heaviside function, \ie~$\grad H_{\Gamma}(\tensor{x}) = \tensor{n} \delta_{\Gamma} (\tensor{x})$. 
Unfortunately, Eq.~\eqref{eq:strain-f} is incompatible with the phase-field setting, because the Heaviside and Dirac delta functions cannot be defined when discontinuous geometry is approximated diffusely.
To overcome this issue, we approximate Eq.~\eqref{eq:strain-f} in the following two ways.
First, we postulate that the displacement jump is constant along the discontinuity, such that $\symgrad\jump{\bm{u}}=\bm{0}$. 
The same postulate was also introduced in the assumed enhanced strain (AES) formulation of Regueiro and Borja~\cite{regueiro2001plane}.
Second, we replace the Dirac delta function by the crack density functional, $\Gamma_d(d, \grad d)$, which converges to the Dirac delta function as $L \rightarrow 0$ ($\Gamma$-convergence).
The same approximation can be found from Verhoosel and de Borst~\cite{verhoosel2013phase}.
As a result, $\tstrain_{f}$ simplifies to
\begin{equation}
	\tstrain_{f} \approx \frac{1}{2}(\jump{\bm{u}} \dyadic \tensor{n} + \tensor{n} \dyadic \jump{\bm{u}}) \Gamma_d(d, \grad d).
  \label{eq:strain-f-approx}
\end{equation}
Hereafter, we shall use Eq.~\eqref{eq:strain-f-approx} to calculate $\tstrain_{f}$ in the phase-field setting.
Further, combining Eqs.~\eqref{eq:disp-decompose-direction} and \eqref{eq:strain-f-approx}, we can calculate the normal closure and tangential slip as
\begin{linenomath}
\begin{align}
	 u_{f} &:= \dfrac{\tstrain_{f}}{\Gamma_d(d,\grad d)}:(\tensor{n} \dyadic \tensor{n}) , \label{eq:u-approx} \\
	v_{f} &:= \dfrac{\tstrain_{f}}{{\Gamma_d(d,\grad d)}}:\tensor{\alpha}, \label{eq:v-approx}
\end{align}
\end{linenomath}
respectively, where
\begin{equation}
  \tensor{\alpha} := (\tensor{n} \dyadic \tensor{m} + \tensor{m} \dyadic \tensor{n})\,
\end{equation}
is the slip direction tensor.

It is also postulated that $\tstrain_{c}$ and $\tstrain_{f}$ can be additively decomposed into elastic and plastic (inelastic) parts as
\begin{linenomath}
\begin{align}
  \tstrain_{c} &= \tstrain^{\el}_{c} + \tstrain^{\pl}_{c} , \\
  \tstrain_{f} &= \tstrain^{\el}_{f} + \tstrain^{\pl}_{f} ,
  \label{eq:strain-decomposition-joint}
\end{align}
\end{linenomath}
where the superscripts $(\cdot)^{\el}$ and $(\cdot)^{\pl}$ denote the elastic and plastic parts, respectively.
Equivalently, the total elastic and plastic strains are
\begin{linenomath}
\begin{align}
  \tstrain^{\el} &= \tstrain^{\el}_{c} + \tstrain^{\el}_{f}, \\
  \tstrain^{\pl} &= \tstrain^{\pl}_{c} + \tstrain^{\pl}_{f},
\end{align}
\end{linenomath}
Although many frictional contact models (\eg~the standard Coulomb friction model) treat frictional slip as entirely inelastic, rough rock fracture can manifest significant elastic deformation due to asperity compressibility and contact effects.
For this reason, elasticity is an integral part of modeling rock fracture with roughness.

Combining Eqs. \eqref{eq:u-approx}, \eqref{eq:v-approx}, and~\eqref{eq:strain-decomposition-joint}, we get the following approximations of the elastic and plastic components of normal and tangential displacement jumps:
\begin{linenomath}
\begin{align}
  u^{\el}_{f} &:= \dfrac{\tstrain^{\el}_{f} : (\tensor{n} \dyadic \tensor{n})}{\Gamma_d(d, \grad d)} ,  \label{eq:u-e} \\
  v^{\el}_{f} &:= \dfrac{\tstrain^{\el}_{f} : \tensor{\alpha}}{\Gamma_d(d, \grad d)} , \label{eq:v-e} \\
  u^{\pl}_{f} &:= \dfrac{\tstrain^{\pl}_{f} : (\tensor{n} \dyadic \tensor{n})}{\Gamma_d(d, \grad d)} ,  \label{eq:u-p} \\
  v^{\pl}_{f} &:=  \dfrac{\tstrain^{\pl}_{f} : \tensor{\alpha}}{\Gamma_d(d, \grad d)} . \label{eq:v-p}
\end{align}
\end{linenomath}

A constitutive model for rock discontinuities can then be introduced to relate the kinematic quantities in Eqs.~\eqref{eq:u-e}--\eqref{eq:v-p} to their corresponding stress components of $\tstress_{f}$.
For this purpose, we also decompose $\tstress_{f}$ into its normal and tangential component as
\begin{linenomath}
\begin{align}
  \stress_{\cn} &:= \tstress_{f}:(\tensor{n} \dyadic \tensor{n}) , \label{eq:normal-stress} \\
  \tau &:= \dfrac{1}{2} \tstress_{f}:\tensor{\alpha} \label{eq:shear-stress},
\end{align}
\end{linenomath}
where $\stress_{\cn}$ is the contact normal stress and $\tau$ is the shear stress in the fracture.
Note that here $\stress_{\cn}$ (and thus the fracture displacement) is considered negative in compression, being consistent with the sign convention in standard phase-field modeling.
In what follows, we formulate a constitutive framework for modeling the compression and shear behavior of rock fractures, from the elastic regime to the plastic regime.
\smallskip

\begin{remark}
  An alternative way to bridge the strain tensor and the displacement jump would be to introduce a characteristic length scale, like how some phase-field models for fracture have calculated crack opening and slip displacements (\eg~\cite{miehe2015minimization,mauthe2017hydraulic,choo2018cracking,bryant2021phase}).
  Yet the new approach proposed in this work---Eqs.~\eqref{eq:strain-decomposition-total}--\eqref{eq:strain-f-approx}---has the following two advantages: (i) it does not introduce any new parameter to the existing phase-field formulation, and (ii) it draws on the $\Gamma$-convergence properties of the crack density functional.
  These two features allow the new phase-field formulation to be mesh-insensitive, like the standard phase-field formulations.
\end{remark}

\subsection{Elastic deformation of rough fracture}
\label{sec:elastic}
We make use of a standard approach to rock joint elasticity that describes the fracture normal and shear response separately.
For the normal traction, we adopt the hyperbolic function proposed by Bandis~\etal~\cite{bandis1983fundamentals}---perhaps the most popular approach in the rock mechanics community---given by
\begin{equation}
  \sigma_{\cn} = \kappa \dfrac{u^{\el}_{f}}{1 - u^{\el}_{f}/u^{\el}_{\max}},
  \label{eq:normal-elastic}
\end{equation}
where $\kappa$ denotes the initial compressive stiffness, and $u^{\el}_{\max}$ the maximum closure of the fracture.
The tangential traction may be described as
\begin{equation}
  \tau = \mu v^{\el}_{f} ,
  \label{eq:tangential-elastic}
\end{equation}
where $\mu$ denotes the shear modulus related to the frictional properties of the fracture.
It is noted that $\mu$ may or may not be dependent on $\sigma_{\cn}$ according to the modeling assumption.

We now reformulate Eqs.~\eqref{eq:normal-elastic} and~\eqref{eq:tangential-elastic} for the phase-field formulation at hand to construct a constitutive relationship between $\tstress_{f}$ and $\tstrain_{f}^{\el}$.
In the present work, we consider the normal and tangential deformations of the fracture and neglect rotations and relative stretching, as in the classic AES method for strong discontinuities~\cite{simo1990class,regueiro2001plane}.
Let us consider the normal deformation first.
Substituting Eq. \eqref{eq:u-e} into Eq. \eqref{eq:normal-elastic} and rearranging the resulting equation gives
\begin{equation}
	\tstrain^{\el}_{f} : (\tensor{n} \dyadic
	\tensor{n}) = \dfrac{\stress_{\cn} \Gamma_d(d, \grad d)}{(\kappa + \stress_{\cn} /u^{\el}_{\max})} .
  \label{eq:normal-eps-sig}
\end{equation}
Inserting Eq.~\eqref{eq:normal-stress} into Eq.~\eqref{eq:normal-eps-sig}, we can rewrite the above equation as
\begin{equation}
  \tstrain^{\el}_{f} : (\tensor{n} \dyadic
  \tensor{n}) = \dfrac{1}{E_{f}} \tstress_{f}:(\tensor{n} \dyadic \tensor{n}) ,
  \label{eq:elastic-normal-strain}
\end{equation}
where $E_{f}$ is the Young's modulus of fracture, defined as
\begin{equation}
	E_{f} := \dfrac{1}{\Gamma_d(d, \grad d)}\left(\kappa + \dfrac{\stress_{\cn}}{u^{\el}_{\max}} \right) .
  \label{eq:E-joint}
\end{equation}
Likewise, for the tangential deformation, we can use Eq.~\eqref{eq:v-e} to rewrite Eq.~\eqref{eq:tangential-elastic} as
\begin{equation}
	\tstrain^{\el}_{f}: \tensor{\alpha} = \dfrac{1}{2G_{f}} \tstress_{f}:\tensor{\alpha} ,
  \label{eq:elastic-tangential-strain}
\end{equation}
where $G_{f}$ is the shear modulus of the fracture, defined as
\begin{equation}
	G_{f} := \dfrac{\mu}{\Gamma_d(d, \grad d)} .
  \label{eq:G-joint}
\end{equation}
To arrive at a constitutive relation between $\tstress_{f}$ and $\tstrain_{f}^{\el}$, we multiply both sides of Eq.~\eqref{eq:elastic-normal-strain} by $\tensor{n} \dyadic \tensor{n}$ and get
\begin{equation}
	\tstrain^{\el}_{f} : [ (\tensor{n} \dyadic
	\tensor{n})\dyadic (\tensor{n} \dyadic
	\tensor{n})] = \dfrac{1}{E_{f}} \tstress_{f}:[ (\tensor{n} \dyadic \tensor{n})\dyadic (\tensor{n} \dyadic \tensor{n})] .
  \label{eq:elastic-normal-strain-normal}
\end{equation}
Similarly, we multiply both sides of Eq.~\eqref{eq:elastic-tangential-strain} by $(1/2)\tensor{\alpha}$ and get
\begin{equation}
	\dfrac{1}{2} \tstrain^{\el}_{f}: (\tensor{\alpha}\dyadic \tensor{\alpha}) = \dfrac{1}{4G_{f}} \tstress_{f}:(\tensor{\alpha}\dyadic \tensor{\alpha}) .
  \label{eq:elastic-tangential-strain-tangential}
\end{equation}
Adding Eqs.~\eqref{eq:elastic-normal-strain-normal} and \eqref{eq:elastic-tangential-strain-tangential}, we obtain the constitutive relation between $\tstress_{f}$ and $\tstrain_{f}^{\el}$ as follows:
\begin{equation}
	\tstrain^{\el}_{f} = \tstress_{f}: \left[\dfrac{1}{E_{f}}(\tensor{n} \dyadic \tensor{n})\dyadic (\tensor{n} \dyadic \tensor{n}) + \dfrac{1}{4G_{f}} \tensor{\alpha}\dyadic \tensor{\alpha} \right] .
  \label{eq:strain-joint-stress-interface}
\end{equation}
The continuous part of the elastic strain, $\tstrain_{c}^{\el}$, is related to $\tstress_{f}$ as
\begin{equation}
	\tstrain^{\el}_{c} = (\mathbb{C}^{\el}_{m})^{-1} : \tstress_{f} .
  \label{eq:strain-solid-stress-interface}
\end{equation}
Since $\tstrain^{\el}=\tstrain^{\el}_{c}+\tstrain^{\el}_{f}$, Eqs.~\eqref{eq:strain-joint-stress-interface} and~\eqref{eq:strain-solid-stress-interface} can be combined as
\begin{equation}
	\tstrain^{\el} = \mathbb{K} : \tstress_{f},
  \label{eq:elastic-strain-interface}
\end{equation}
where
\begin{equation}
  \mathbb{K} := (\mathbb{C}^{\el}_{m})^{-1} + \dfrac{1}{E_{f}} \left(\tensor{n} \dyadic\tensor{n} \right) \dyadic\left(\tensor{n} \dyadic\tensor{n} \right)  + \dfrac{1}{4G_{f}} \left(\tensor{\alpha} \dyadic\tensor{\alpha} \right) .
  \label{eq:elastic-compliance}
\end{equation}
Note that $\mathbb{K}$ is not constant and depends on the stress state. 
This equation then allows us to derive the elastic tangent of the fracture stress tensor as
\begin{equation}
	\mathbb{C}^{\el}_{f} = \pdv{\tstress_{f}}{ \tstrain^{\el}}   .
\end{equation}
Noting that $E_{f}$ and $G_{f}$ depend on $\tstress_{f}$ via Eqs.~\eqref{eq:E-joint} and \eqref{eq:G-joint}, we take derivatives of both sides of Eq.~\eqref{eq:elastic-strain-interface} with respect to $\tstrain^{\el}$ and obtain
\begin{linenomath}
\begin{align}
	\mathbb{I} &= \mathbb{K} :\pdv{ \tstress_{f}}{ \tstrain^{\el}} + \pdv{ \mathbb{K}}{ \tstrain^{\el}}:\tstress_{f} \\
	 & = \mathbb{K} : \mathbb{C}^{\el}_{f} + \left(\pdv{ \mathbb{K}}{ \tstress_{f}}:\mathbb{C}^{\el}_{f} \right):\tstress_{f} ,
   \label{eq:elastic-strain-interface-deriv}
\end{align}
\end{linenomath}
Rearranging Eq.~\eqref{eq:elastic-strain-interface-deriv} gives $\mathbb{C}^{\el}_{f}$ as
\begin{linenomath}
\begin{align}
	\mathbb{C}^{\el}_{f} =  \left[\mathbb{K} + \left(\pdv{ \mathbb{K}}{ \tstress_{f}}:\tstress_{f} \right) \right]^{-1} ,
  \label{eq:elastic-cto-interface}
\end{align}
\end{linenomath}
where
\begin{linenomath}
\begin{align}
	\pdv{ \mathbb{K}}{ \tstress_{f}}:\tstress_{f} &= \left(\pdv{ \mathbb{K}}{ \stress_{\cn}} \dyadic \pdv{ \stress_{\cn}}{ \tstress_{f}} \right) : \tstress_{f} \nonumber\\
	&= - \dfrac{1}{\Gamma_d(d, \grad d) E^2_{f}}\dfrac{\stress_{\cn}}{u^{\el}_{\max}} \left(\tensor{n} \dyadic\tensor{n} \right) \dyadic\left(\tensor{n} \dyadic\tensor{n} \right) - \dfrac{1}{4\Gamma_d(d, \grad d)G^2_{f}} \stress_{\cn} \mu'(\stress_{\cn}) \tensor{\alpha} \dyadic \tensor{\alpha} , \label{eq:dK-dsigma}
\end{align}
\end{linenomath}
with $\mu'(\stress_{\cn})$ denoting the derivative of $\mu$ with respect to $\stress_{\cn}$.
Note that $\mu'(\stress_{\cn})=0$ if $\mu$ is considered stress-independent.

\subsection{Inelastic deformation of rough fracture}
\label{sec:inelastic}
For modeling the inelastic deformation of rough rock fractures, we adopt a plasticity-like framework, which is standard for physically-motivated models for rock joints~(\eg~\cite{son2004elasto,white2014anisotropic}).
The general form of the yield function may be written as
\begin{equation}
	F = \tau + \sigma_{\cn} \tan(\phi_\mathrm{b} + \omega\,\psi) \leq 0,
  \label{eq:yield-function}
\end{equation}
where $\phi_\mathrm{b}$ is the basic friction angle, $\psi$ the dilation angle, and $\omega$ is an abrasion coefficient which accounts for the impact of asperity damage on mobilized shear strength. 
% \todo{I introduced ``abrasion'' to avoid ambiguity with phase field ``damage'', and because it may be confusing that $\omega > 1$ when talking about a damage coefficient.}
Here we consider $\phi_\mathrm{b}$ a constant, while viewing $\psi$ and $\omega$ as state variables.
Next, we consider the potential function of the following general form~\cite{son2004elasto}
\begin{equation}
	G = \tau + \int \tan(\psi)\, \dd\sigma_{\cn},
  \label{eq:potential-function}
\end{equation}
which gives the non-associative flow rule as
\begin{linenomath}
\begin{align}
  \dot{\tstrain}^{\pl}_{f} = \dot{\lambda}\pdv{ G}{ \tstress_{f}}
  &= \dot{\lambda} \left(\pdv{G}{\tau}\pdv{\tau}{\tstress_{f}} + \pdv{G}{\stress_{\cn}}\pdv{\stress_{\cn}}{\tstress_{f}}  \right) \nonumber\\
  &= \dot{\lambda}\left[\dfrac{1}{2}\tensor{\alpha} + \tan \psi (\tensor{n} \dyadic\tensor{n})\right] ,
  \label{eq:flow-rule}
\end{align}
\end{linenomath}
with $\lambda$ denoting the plastic multiplier.
It is noted that the form of potential function~\eqref{eq:potential-function} accommodates a stress-dependent dilation angle, while keeping the definition of the dilation angle as $\tan\psi = \dd u^{\pl}_{f}/ \dd v^{\pl}_{f}$.

We now specialize the general framework to the constitutive model proposed by White~\cite{white2014anisotropic} to match experimentally observed behavior of rough fractures.
Under monotonic (non-cyclic) loading, the dilation is given by
\begin{equation}
  \tan \psi = \tan \psi_\mathrm{r} \left[\tanh \left(\dfrac{v^{\pl}_{f}}{b} \right) \right], \label{eq:white-dilation}
\end{equation}
where $\psi_\mathrm{r}$ denotes a residual dilation angle, and $b$ is a characteristic slip length.
In the seated position of the fracture ($v^p_f=0$) the initial dilation angle is zero. With accumulating slip ($v^p_f > 0$) asperities ride up over one another and the dilation grows towards a residual dilation angle.
The abrasion coefficient controls the mobilized shear strength via Eq.~(\ref{eq:yield-function}), and is given by
\begin{equation}
  \omega = 1 + (\omega_{0} - 1)\exp(-c\Lambda), \label{eq:white-damage}
\end{equation}
where $\omega_{0}$ is an abrasion parameter controlling the peak shear strength, $c$ is a softening constant, and $\Lambda$ denotes a history variable that quantifies the degradation of roughness.
When the abrasion is assumed isotropic in all directions, it can be equated to the cumulative plastic slip, \ie
\begin{equation}
	\Lambda = \int_{0}^{T} \dot{v}^{\pl}_{f} \: \dd t . \label{eq:white-Lambda}
\end{equation}
It is noted that in the phase-field formulation, one can calculate $\dot{v}^{\pl}_{f}$ by taking time derivatives on both sides of Eq.~\eqref{eq:v-p}. This model allows for an initial peak in strength that then degrades as slip accumulates and asperities are worn away, as often seen in rough fracture experiments. See White~\cite{white2014anisotropic} for further details.

% SECTION
% ------------------------------------------------------------------------------
\section{Discretization and algorithm}
\label{sec:discretization}
This section describes how to numerically solve the proposed phase-field formulation.

\subsection{Finite element discretization}
The standard continuous-Galerkin finite element method can be used to solve the two-field governing equations, Eqs.~\eqref{eq:momentum-balance} and~\eqref{eq:phase-field-evolution}, in which the unknown fields are the displacement vector field $\tensor{u}$ and the phase field $d$.
We introduce spaces for the trial solutions as
\begin{linenomath}
\begin{align}
	\mathcal{S}_{u} &:= \{\tensor{u} \, \rvert \, \tensor{u} \in H^{1} , \,\,  \tensor{u} = \hat{\tensor{u}} \,\, \text{on} \,\, \pd_{u} \Omega   \} , \\
	\mathcal{S}_{d} &:= \{d \, \rvert\, d \in H^{1} \} ,
\end{align}
\end{linenomath}
where $H^{1}$ denotes a Sobolev space of order one.
Accordingly, spaces for the weighting functions are defined as
\begin{linenomath}
\begin{align}
	\mathcal{V}_{u} &:= \{ \tensor{\eta} \, \rvert \, \tensor{\eta} \in H^{1} , \, \, \tensor{\eta} = \tensor{0} \, \, \text{on} \, \, \pd_{u} \Omega \}, \\
	\mathcal{V}_{d} &:= \{\phi \, \rvert \, \phi \in H^{1} \} .
\end{align}
\end{linenomath}
Through the standard weighted residual procedure, we arrive at the variational form of the governing equations, 
\begin{linenomath}
\begin{align}
  \tensor{\mathcal{R}}_{u} &:= - \int_{\Omega} \symgrad \tensor{\eta}:\tstress \: \dd V + \int_{\Omega} \tensor{\eta} \cdot \rho \tensor{g} \: \dd V + \int_{\pd_{t} \Omega} \tensor{\eta} \cdot \hat{\tensor{t}} \: \dd A  = 0 ,
  \label{eq:momentum-balance-residual} \\
	\tensor{\mathcal{R}}_{d} &:= \int_{\Omega} \phi g'(d) \mathcal{H}^{+} \: \dd V + \int_{\Omega} \dfrac{3 \mathcal{G}_{c}}{8L}(2L^2 \grad \phi \cdot \grad d + \phi) \: \dd V = 0 .
  \label{eq:pf-evolution-residual}
\end{align}
\end{linenomath}
The finite element discretization of Eqs.~\eqref{eq:momentum-balance-residual} and \eqref{eq:pf-evolution-residual} is standard and its details are skipped for brevity.
When modeling non-propagating rock fractures, Eq.~\eqref{eq:pf-evolution-residual} needs to be solved only once for the initialization of the phase field.
To simulate propagating fractures, however, both Eqs.~\eqref{eq:momentum-balance-residual} and \eqref{eq:pf-evolution-residual} should be solved in every load step.
In either case, Eq.~\eqref{eq:pf-evolution-residual} can be solved separately in the same manner as existing phase-field models, because it is common to solve the two governing equations in a staggered manner~\cite{miehe2010phase}.
Therefore, in what follows, we focus on the solution of Eq.~\eqref{eq:momentum-balance-residual}.

Given that the constitutive relation is nonlinear, we use Newton's method to solve the problem at hand.
At each Newton iteration, we solve the Jacobian system given by
\begin{equation}
	\tensor{\mathcal{J}}_{u} \Delta \tensor{U} = -\tensor{\mathcal{R}}_{u}^{h} \,,
  \label{eq:newton-system}
\end{equation}
where $\tensor{\mathcal{J}}_{u}$ denotes the Jacobian matrix, $\Delta\tensor{U}$ the nodal displacement increment, and $\tensor{\mathcal{R}}_{u}^{h}$ the discretized version of Eq.~\eqref{eq:momentum-balance-residual}.
The specific expression of $\tensor{\mathcal{J}}_{u}$ is
\begin{equation}
	\tensor{\mathcal{J}}_{u} := - \int_{\Omega} \symgrad \tensor{\eta}^{h} : \mathbb{C} : \symgrad \tensor{\eta}^{h} \: \dd V , \label{eq:jacobian-disp-discrete}
\end{equation}
where $\tensor{\eta}^{h}$ is the discretized version of $\bm{\eta}$, and $\mathbb{C}$ denotes the stress-strain tangent operator.
To evaluate Eq.~\eqref{eq:newton-system} at each Newton iteration, one needs to update the stress tensor and the tangent operator at every material (quadrature) point.
An algorithm for updating these two quantities is designed below.

\subsection{Material update algorithm}
Algorithm \ref{algo:material-update} presents a detailed procedure to update the stress tensor and tangent operator at every material point.
Quantities at the previous time step $t_n$ are denoted with subscript $(\cdot)_{n}$, whereas quantities at the current time step $t_{n+1}$ are written without any subscript to avoid proliferation of subscripts.
The algorithm is essentially the same as the predictor--corrector algorithm in the phase-field method for frictional interfaces~\cite{fei2020phasea}, except the following three modifications.
First, we have introduced a boolean flag \texttt{open\underline{ }flag} to keep the material point in the open mode when its contact condition is identified to be open in a Newton iteration. 
In every load step, the flag is initialized as false at the beginning and switched to true if the contact condition becomes open during a Newton iteration.
We have experienced that the use of this flag makes the Newton iteration more robust.
Second, due to the nonlinearity of elastic fracture deformation, we have designed a local Newton iteration to evaluate the interface stress tensor $\tstress_{f}$ and the interface elastic tangent operator $\mathbb{C}^{\el}_{f}$---see Algorithm \ref{algo:interface-stress-update}.
Third, we evaluate the yield function in a semi-implicit manner, using the normal stress in the previous step. This semi-implicit approach greatly simplifies the calculation of the derivative of the yield function and thus the inelastic fracture tangent operator $\mathbb{C}_{f}$.
Lastly, we have added a small tolerance $k_{\text{tol}}$ (\eg~$10^{-3}$) to the weight of the bulk stress tensor in the calculation of the overall stress tensor (\cf~Eq.~\eqref{eq:stress-partition}), to avoid ill-posedness of the matrix system~\cite{miehe2010thermodynamically}.
Although such a tolerance is often found to be unnecessary in standard phase-field models, we have found that its use is critical to numerical robustness of the phase-field formulation at hand, particularly for intersecting cracks.
\begin{algorithm}[h!]
    \setstretch{1.25}
    \caption{Material point update procedure for the phase-field model for rough rock fractures}
      \begin{algorithmic}[1]
      	\Require $\Delta \tstrain$, $d$, $\tensor{n}$ and $\tensor{m}$ at $t_{n+1}$.
      	\State Calculate $\tstrain = \tstrain_{n} + \Delta \tstrain$, and $\tstress_{m} = \mathbb{C}^{\el}_{m}:\tstrain$.
      	\If {$d = 0$}
      		\State Intact region.
      		Return $\tstress = \tstress_{m}$, and $\mathbb{C}= \mathbb{C}_{m}^{\el}$.
      	\EndIf
      	\State Calculate the normal strain at the interface region, $\strain_{\cn} = \tstrain:(\tensor{n} \dyadic \tensor{n})$.
      	\If {$\strain_{\cn} > 0$ or \texttt{open\underline{ }flag}}
      		\State Open state.
      		 \texttt{open\underline{ }flag} = true.
      		\State Return $\tstress = g(d) \tstress_{m}$, and $\mathbb{C} = g(d)\mathbb{C}^{\el}_{m}$.
      	\EndIf
      	\State Closed state.
      	\State Calculate the trial elastic strain, $\tstrain^{\el,\tr} = \tstrain^{\el,\tr}_{n} + \Delta \tstrain$.
      	\State Update the trial stress and the elastic tangent operator in the fracture region ($\tstress^{\tr}_{f}$ and $\mathbb{C}^{\el}_{f}$) using Algorithm \ref{algo:interface-stress-update}.
      	\State Update the dilation angle $\psi$ and the damage coefficient $\omega$.
      	\State Calculate the yield function $F = \tau^{\tr} + \stress_{\cn,n} \tan(\phi_\mathrm{b} + \omega_{n}\psi_{n})$, where $\tau^{\tr} := 1/2 \tstress^{\tr}_{f} : \tensor{\alpha}$ and $\stress_{\cn,n} := \tstress_{f,n}:(\tensor{n} \dyadic \tensor{n})$.
      	\If {$F < 0$}
      		\State Elastic fracture deformation.
      		Update $\tstrain^{\el} = \tstrain^{\el,\tr}$, $\tstrain^{\pl} = \tstrain^{\pl}_{n}$, $\tstress_{f} = \tstress^{\tr}_{f}$, and $\mathbb{C}_{f} = \mathbb{C}^{\el}_{f}$.
      	\Else
      		\State Inelastic fracture deformation.
      		Perform return mapping to update $\tstress_{f}$, $\mathbb{C}_{f}$,  $\tstrain^{\el}$, $\psi$, and $\omega$.
      	\EndIf
      	\State Update $\tstress = [g(d) + k_\mathrm{tol} ] \tstress_{m} + [1 - g(d)] \tstress_{f}$, and $\mathbb{C} = [g(d) + k_\mathrm{tol} ] \mathbb{C}^{\el}_{m} + [1 - g(d)] \mathbb{C}_{f}$.
      	\Ensure $\tstress$ and $\mathbb{C}$ at $t_{n+1}$.
      \end{algorithmic}
  \label{algo:material-update}
\end{algorithm}

\begin{algorithm}[h!]
    \setstretch{1.25}
    \caption{Update procedure for the fracture stress and the fracture elastic tangent operator}
      \begin{algorithmic}[1]
      	\Require $\tstrain^{\el}$, $\tensor{n}$ and $\tensor{m}$.
      	  \State Set the iteration counter $k=0$.
          \State Initialize the fracture stress tensor $\tstress^{k}_{f} = \tstress_{f,n}$, and the elastic compliance tensor $\mathbb{K}^{k} = \mathbb{K}_{n}$.
          \Repeat
      		\State $k = k + 1$.
      		\State Let $\tstress^{k} = \tstress^{k-1}_{f}$, and $\mathbb{K}^{k} = \mathbb{K}^{k-1}$.
      		\State Calculate $\tensor{r}^{k} = \mathbb{K}^{k} : \tstress^{k}_{f} - \tstrain^{\el}$, and $\tensor{J}^{k} = \mathbb{K}^{k} - (\pd \mathbb{K}/\pd \tstress_{f})^{k} : \tstress^{k}_{f}$, according to Eq.~\eqref{eq:dK-dsigma}.
      		\State Calculate $\Delta \tstress^{k}_{f}$ by solving $\tensor{J}^{k} : \Delta \tstress^{k}_{f} = -\tensor{r}^{k}$.
      		\State Update $\tstress^{k}_{f} = \tstress^{k - 1}_{f} + \Delta \tstress^{k}_{f}$.
      		\State Calculate $\stress_{\cn} = \tstress^{k}_{f}:(\tensor{n} \dyadic \tensor{n})$.
      		\State Update $E_{f}$ and $G_{f}$ according to Eqs.~\eqref{eq:E-joint} and \eqref{eq:G-joint}.
          \State Update $\mathbb{K}^{k}$ according to Eq.~\eqref{eq:elastic-compliance}.
      	\Until {$\| \tensor{r}^{k} \| < \text{tol}$}
      	\State Update $\tstress_{f} = \tstress^{k}_{f}$
        \State Calculate $\mathbb{C}^{\el}_{f}$ according to Eq.~\eqref{eq:elastic-cto-interface}, with $\mathbb{K}^{k}$.
      	\Ensure $\tstress_{f}$ and $\mathbb{C}^{\el}_{f}$.
      \end{algorithmic}
  \label{algo:interface-stress-update}
\end{algorithm}

\subsection{Return mapping and inelastic tangent operator}
\label{sec:return-mapping}
In Algorithm \ref{algo:material-update}, when the trial stress violates the requirement $F\leq 0$, we use a return mapping algorithm to correct the trial stress and strains such that they satisfy $F=0$.
The return mapping is performed in the full strain space and described in the following.

The unknowns in the return mapping are the six independent components of the strain tensor and the discrete plastic multiplier. Using Kelvin notation for an elegant representation of tensor algebra~\cite{nagel2016advantages}, we write the unknown vector as
\begin{equation}
  \tensor{x} =
  \begin{bmatrix}
    \left(\tstrain^{\el} \right)_{6 \times 1} \\
    \Delta \lambda
  \end{bmatrix}_{7 \times 1}.
\end{equation}
The equations to be satisfied can be written as residuals
\begin{equation}
  \tensor{r} =
  \begin{bmatrix}
    \left(\tstrain^{\el} - \tstrain^{\el,\tr} + \Delta \lambda \dfrac{\pd G}{\pd \tstress_{f}}  \right)_{6 \times 1} \\
    F
  \end{bmatrix}_{7 \times 1} \rightarrow 0.
  \label{eq:return-mapping-residual}
\end{equation}
It is again noted that we evaluate the value of $F$ in the residual vector~\eqref{eq:return-mapping-residual} in a semi-implicit way, using $\stress_{\cn,n} := \tstress_{f,n}:(\tensor{n} \dyadic \tensor{n})$.

Then we use Newton's method to find a numerical solution.
At each Newton iteration, we solve
\begin{equation}
	\tensor{J} \cdot \Delta\tensor{x} = -\tensor{r},
  \label{eq:return-mapping-linearized}
\end{equation}
for the search direction $\Delta\bm{x}$.
The Jacobian matrix is given by
\begin{equation}
	\tensor{J} =
	\begin{bmatrix}
		\left(\mathbb{I} + \Delta \lambda \dfrac{\pd^2 G}{\pd \tstress_{f} \dyadic \pd \tstress_{f}}:\mathbb{C}^{\el}_{f}  \right)_{6 \times 6} & \left( \dfrac{\pd G}{\pd \tstress_{f}} + \Delta \lambda \dfrac{\pd^2 G}{\pd \tstress_{f} \pd \Delta \lambda} \right)_{6 \times 1} \\
		\\
		\left(\dfrac{\pd F}{\pd \tstress_{f}}:\mathbb{C}^{\el}_{f} \right)^{\intercal}_{1 \times 6}  & \stress_{\cn, n} \dfrac{\pd \tan(\phi_{b} + \omega \psi)}{\pd \Delta \lambda}
	\end{bmatrix}_{7 \times 7} . \label{eq:return-mapping-jacobian}
\end{equation}
The derivatives required to complete the Jacobian matrix are provided in~\ref{appendix:derivates}.
It is noted that the tensorial operations above have been written in Kelvin notation. 

Once the return mapping is completed, we calculate the inelastic fracture tangent operator defined as
\begin{equation}
  \mathbb{C}_{f} = \dfrac{\pd \tstress_{f}}{\pd \tstrain^{\el, \tr}} \, .
\end{equation}
To evaluate this inelastic tangent operator, here we adopt the method used in Bryant and Sun~\cite{bryant2021phase}.
To begin, we consider the trial elastic strain as a variable and re-linearize Eq.~\eqref{eq:return-mapping-residual} following Eq.~(7.127) in de Souza Neto~\etal~\cite{de2011computational}.
This gives
\begin{equation}
  \begin{bmatrix}
    \left(\dd \tstrain^{\el} + \Delta \lambda \dfrac{\pd^2 G}{\pd \tstress_{f} \dyadic \pd \tstress_{f}}:\mathbb{C}^{\el}_{f}:\dd \tstrain^{\el} + \dfrac{\pd G}{\pd \tstress_{f}} \dd \Delta \lambda + \Delta \lambda \dfrac{\pd^2 G}{\pd \tstress_{f} \pd \Delta \lambda} \dd \Delta \lambda \right)_{6 \times 1} \\
    \dfrac{\pd F}{\pd \tstress_{f}}:\mathbb{C}^{\el}_{f}:\dd \tstrain^{\el} +  \stress_{\cn,n} \dfrac{\pd \tan(\phi_{b} + \omega \psi)}{\pd \Delta \lambda} \dd \Delta \lambda
  \end{bmatrix}_{7 \times 1}
  =
  \begin{bmatrix}
    \left(\dd \tstrain^{\el,\tr} \right)_{6 \times 1} \\
    0
  \end{bmatrix}_{7 \times 1} . \label{eq:return-mapping-linearized-cto}
\end{equation}
We then insert $\dd \tstrain^{\el} = (\mathbb{C}^{\el}_{f})^{-1}:\dd \tstress_{f}$ into the above equation and obtain
\begin{linenomath}
\begin{align}
  \left[({\mathbb{C}^{\el}}_{f})^{-1} + \Delta \lambda \dfrac{\pd^2 G}{\pd \tstress_{f} \dyadic \pd \tstress_{f}} \right] : \dd \tstress_{f} + \left( \dfrac{\pd G}{\pd \tstress_{f}} + \Delta \lambda \dfrac{\pd^2 G}{\pd \tstress_{f} \pd \Delta \lambda} \right) \dd \Delta \lambda &= \dd \tstrain^{\el,\tr} ,\label{eq:d-strain-trial-eq} \\
   \dfrac{\pd F}{\pd \tstress_{f}} : \dd \tstress_{f} + \stress_{\cn,n} \frac{\pd \tan(\phi_{b} + \omega \psi)}{\pd \Delta \lambda} \dd \Delta \lambda  &= 0 . \label{eq:d-lambda-eq}
\end{align}
\end{linenomath}
Rearranging Eq.~\eqref{eq:d-strain-trial-eq} gives~\cite{cuitino1992material,de1994note}
\begin{equation}
	\dd \tstress_{f} = {\mathbb{P}} : \left[\dd \tstrain^{\el,\tr} -  \left( \dfrac{\pd G}{\pd \tstress_{f}} + \Delta \lambda \dfrac{\pd^2 G}{\pd \tstress_{f} \pd \Delta \lambda} \right) \dd \Delta \lambda \right] , \label{eq:d-stress-d-strain-tmp}
\end{equation}
where
\begin{equation}
	{\mathbb{P}} = \left(\mathbb{I} + \mathbb{C}^{\el}_{f} \Delta \lambda \dfrac{\pd^2 G}{\pd \tstress_{f} \dyadic \pd \tstress_{f}} \right)^{-1}:\mathbb{C}^{\el}_{f} .
\end{equation}
Differentiating Eq.~\eqref{eq:d-stress-d-strain-tmp} with respect to the trial elastic strain gives the inelastic interface tangent operator as
\begin{equation}
	\mathbb{C}_{f} = \dfrac{\dd \tstress_{f}}{\dd \tstrain^{\el , \tr}} =  {\mathbb{P}} : \left[\mathbb{I} -  \left( \dfrac{\pd G}{\pd \tstress_{f}} + \Delta \lambda \dfrac{\pd^2 G}{\pd \tstress_{f} \pd \Delta \lambda} \right) \otimes \dfrac{\dd \Delta \lambda}{\dd \tstrain^{\el, \tr}} \right]. \label{eq:inelastic-cto}
\end{equation}
The last term in the bracket, \ie~$\dd \Delta \lambda /\dd \tstrain^{\el, \tr}$, can be calculated by substituting Eq.~\eqref{eq:d-stress-d-strain-tmp} into Eq.~\eqref{eq:d-lambda-eq}.
This operation gives
\begin{equation}
  \dfrac{\dd \Delta \lambda}{\dd \tstrain^{\el, \tr}} = \dfrac{\mathbb{P}: \dfrac{\pd F}{\pd \tstress_{f}}}{\left( \dfrac{\pd G}{\pd \tstress_{f}} + \Delta \lambda \dfrac{\pd^2 G}{\pd \tstress_{f} \pd \Delta \lambda} \right) : \left(\mathbb{P}: \dfrac{\pd F}{\pd \tstress_{f}} \right) -  \stress_{\cn,n}} \dfrac{\pd \tan(\phi_{b} + \omega \psi)}{\pd \Delta \lambda}.
	% \tensor{A}:\dd \tstrain^{\el,\tr} = \left[ \tensor{B}:\tensor{A} -  \stress_{\cn} \frac{\pd \tan(\phi_{b} + \omega \psi)}{\pd \Delta \lambda^{\pl}} \right] \dd \Delta \lambda^{\pl} ,
\end{equation}
Inserting the above equation into Eq.~\eqref{eq:inelastic-cto} gives the final form of the inelastic fracture tangent operator as follows:
\begin{equation}
	\mathbb{C}_{f} =  \mathbb{P} -  \dfrac{ \left[\mathbb{P} : \left( \dfrac{\pd G}{\pd \tstress_{f}} + \Delta \lambda \dfrac{\pd^2 G}{\pd \tstress_{f} \pd \Delta \lambda} \right) \right] \otimes \left(\mathbb{P}: \dfrac{\pd F}{\pd \tstress_{f}}\right) }{\left( \dfrac{\pd G}{\pd \tstress_{f}} + \Delta \lambda \dfrac{\pd^2 G}{\pd \tstress_{f} \pd \Delta \lambda} \right) : \left(\mathbb{P}: \dfrac{\pd F}{\pd \tstress_{f}} \right) -  \stress_{\cn,n} \dfrac{\pd \tan(\phi_{b} + \omega \psi)}{\pd \Delta \lambda}}. \label{eq:inelastic-cto-final}
\end{equation}
\smallskip

\begin{remark}
  The algorithm presented above can be used for both stationary and evolving fractures, provided that the two governing equations are solved in a staggered way. Note that such a staggered solution scheme has been commonly employed in the vast majority of phase-field fracture simulations. 
\end{remark}

% SECTION
% ------------------------------------------------------------------------------
\section{Numerical examples}
\label{sec:examples}

In this section, we first verify the proposed phase-field formulation by comparing its numerical results with those obtained by a well-established method for discrete modeling of discontinuities, namely the extended finite element method (XFEM)~\cite{moes1999finite}.
The contact treatment of our XFEM employs the algorithm proposed by Liu and Borja~\cite{liu2008contact}.
For a thorough verification, we compare the results of the two methods in a variety of problems, from shearing of a single discontinuity to compression of fractured rocks with a single crack, two non-intersecting cracks, and two intersecting cracks.
Following the verification, we extend the last two examples (two non-intersecting cracks and two intersecting cracks) to propagating fractures, to demonstrate the capabilities of the phase-field formulation for simulating complex crack growth from rough discontinuities.

We use the same set of material parameters for all the numerical examples.
The parameters of the fracture model are adopted from White~\cite{white2014anisotropic}, which was calibrated against the experimental data of Wibowo \etal~\cite{wibowo1994effect}.
See Table \ref{tab:white-parameters} for the parameters.
Regarding the elasticity parameters of the bulk matrix, a bulk modulus of $K = 16.67$ GPa and a shear modulus of $G = 12.5$ GPa are assigned.
\begin{table}[h!]
    \centering
    \begin{tabular}{l|l|l|c}
    \toprule
    Parameter & Symbol & Unit & Value  \\
    \midrule
    Maximum joint closure & $u^{\el}_{\max}$ & mm & -0.2\\
    Initial compressive modulus & $\kappa$ & MPa/mm & 11.0 \\
    Joint shear modulus & $\mu$ & MPa/mm & 20.0 \\
    Basic friction angle & $\phi_\mathrm{b}$ & degrees & 27.4 \\
    Residual dilation angle & $\psi_\mathrm{r}$ & degrees & 6.1 \\
    Characteristic slip & $b$ & mm & 0.5 \\
    Abrasion coefficient & $\omega_{0}$ & - & 3.3 \\
    Softening coefficient & $c$ & - & 0.15 \\
    \bottomrule
    \end{tabular}
    \caption{Material parameters of the fracture model, calibrated against the experimental data of Wibowo \etal~\cite{wibowo1994effect}. See White~\cite{white2014anisotropic} for details of the calibration procedure.}
    \label{tab:white-parameters}
\end{table}

All the problems in this section are prepared through several common protocols described in the following.
We first discretize the domain using a regularly structured mesh with monosized quadrilateral elements.
To represent the discontinuities, we initialize the phase field by solving Eq.~\eqref{eq:pf-evolution-residual} with prescribed $\mathcal{H}^{+}$, which is a standard approach in the community~\cite{borden2012phase,fei2020phasea}.
We then locally refine the elements where $d > 0$ according to a prescribed value of $L/h$.
We neglect body forces, assume plane-strain conditions, and use linear shape functions.
We evaluate the shear stress and dilation of the cracks at quadrature points closest to the discontinuities.
We obtain the numerical results from our in-house finite element code \texttt{Geocentric}, which is built on the open source finite element library \texttt{deal.II}~\cite{arndt2021deal} and has been used in a number of previous studies (\eg~\cite{choo2016hydromechanical,white2016block,choo2018large,choo2019stabilized}).

\subsection{Shearing of a single discontinuity}
\begin{figure}[h!]
  \centering
  \includegraphics[width=0.5\textwidth]{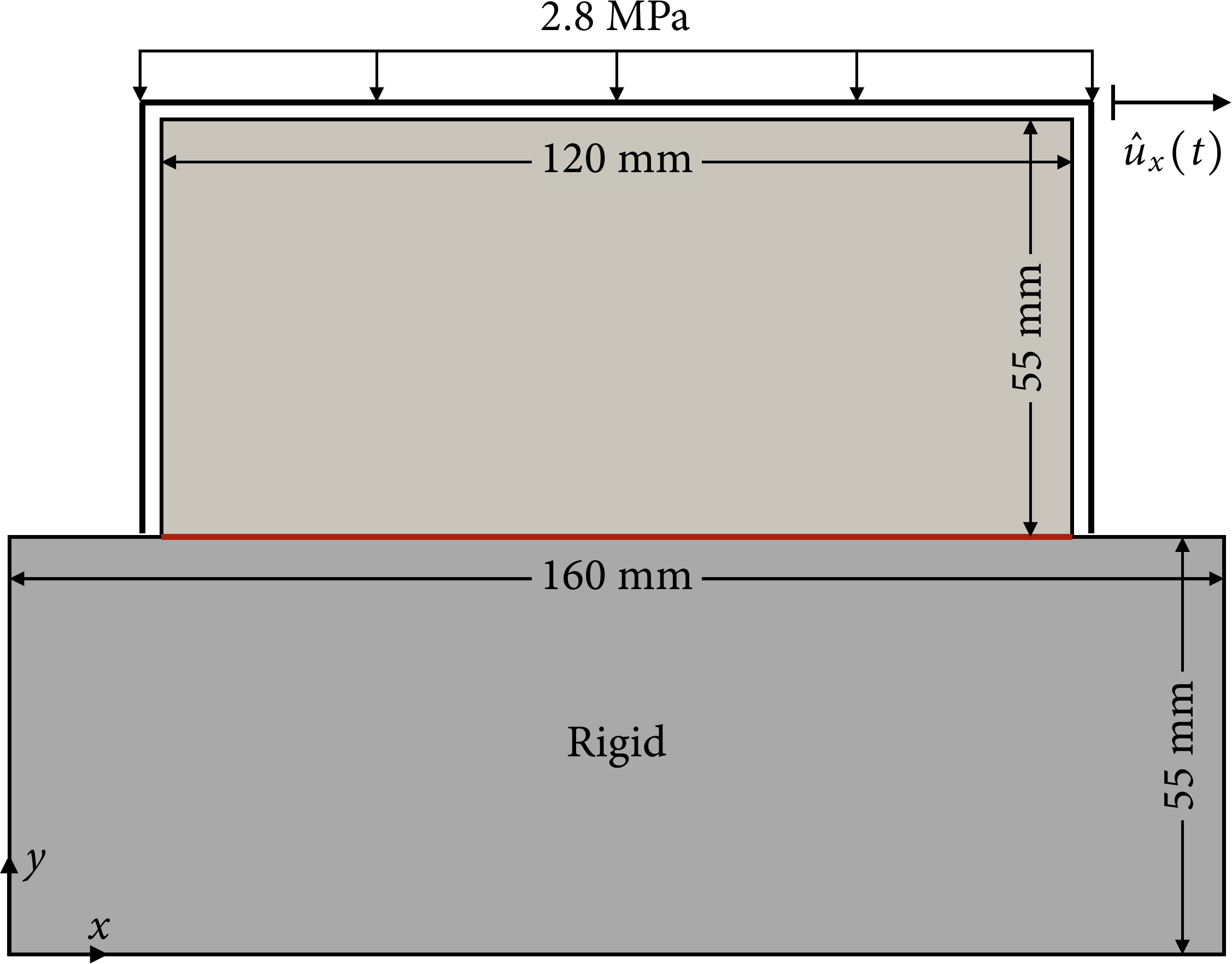}
  \caption{Shearing of a single discontinuity: geometry and boundary conditions.}
  \label{fig:shearing-setup}
\end{figure}

\begin{figure}[h!]
  \centering
  \includegraphics[width=1.0\textwidth]{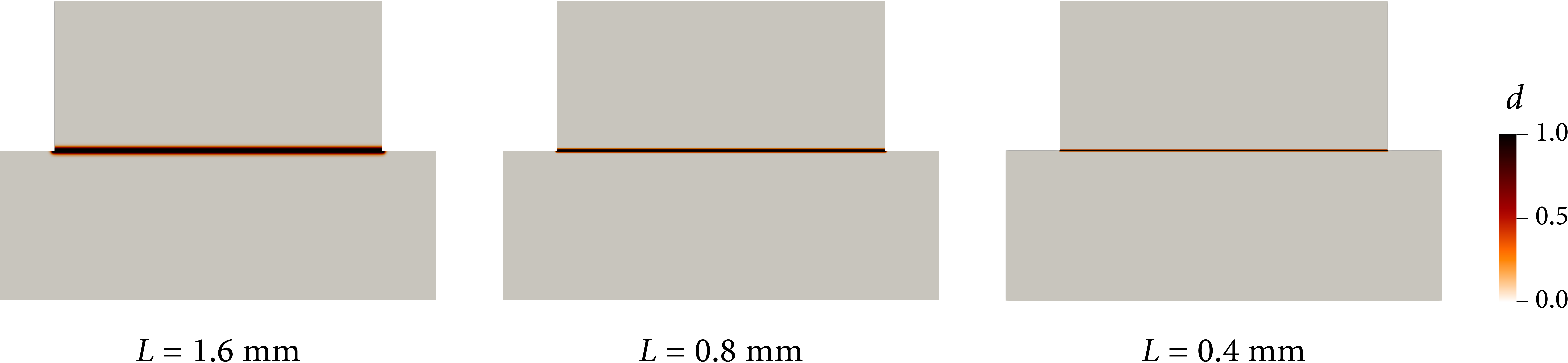}
  \caption{Shearing of a single discontinuity: phase-field distributions initialized with three different phase-field length parameters.}
  \label{fig:shearing-pf}
\end{figure}

We begin by simulating shearing of a single discontinuity, which is the most straightforward setting for studying the behavior of a rough rock fracture.
We adopt the configuration of a shear test performed by Lee~\etal~\cite{lee2001influence}, in which an elastic block under a constant normal stress slides along a wider rigid block fixed to the bottom boundary.
The specific geometry and boundary conditions are illustrated in Fig.~\ref{fig:shearing-setup}.
To investigate the sensitivity of the numerical results to the phase-field length parameter, we prepare three different phase-field distributions initialized with $L=1.6$ mm, $0.8$ mm and $0.4$ mm, as shown in Fig.~\ref{fig:shearing-pf}.
We then apply a prescribed horizontal displacement on the boundary of the upper block, 
except for the region where $d>0$ (\ie~the diffusely approximated discontinuity) for which a displacement boundary condition cannot be imposed (see Bryant and Sun~\cite{bryant2021phase} for a similar treatment).
The rigid block is meshed but its displacement degrees of freedom are fixed.
The XFEM solution to this problem is obtained by embedding the discontinuity into a rectangular domain, similar to how other XFEM studies have modeled similar problems (\eg~\cite{annavarapu2014nitsche,choo2021barrier}).
The simulation proceeds with a uniform displacement rate of $0.1$ mm until the total horizontal displacement reaches $10$ mm ($100$ steps in total).

\begin{figure}[h!]
    \centering
    \subfloat[Shear stress]{\includegraphics[height=0.4\textwidth]{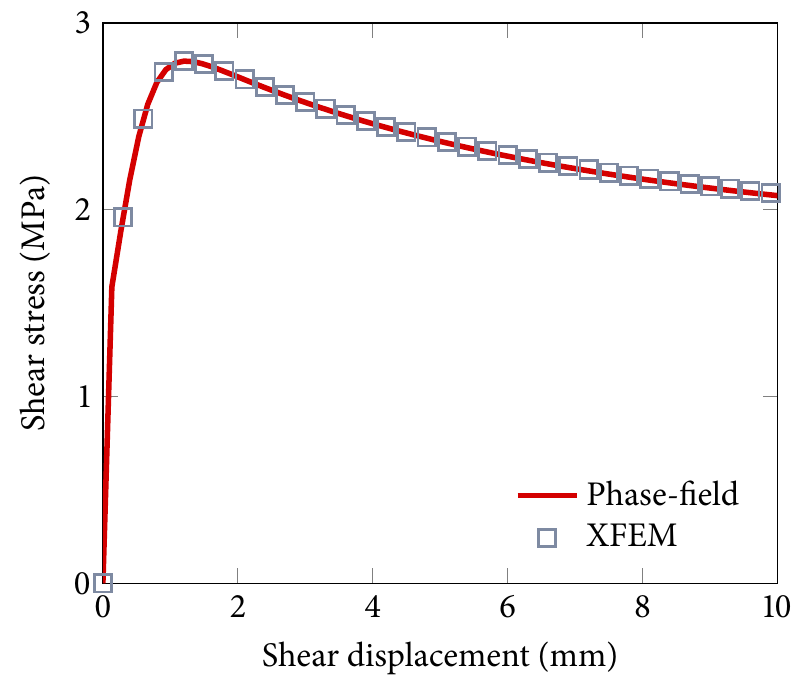}} \hspace{0.5em}
    \subfloat[Dilation]{\includegraphics[height=0.4\textwidth]{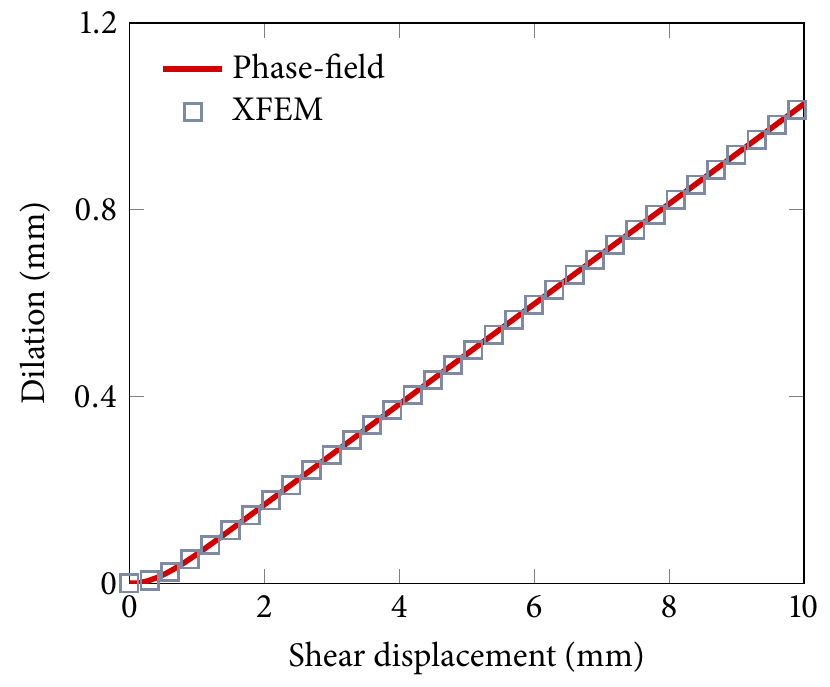}}
    \caption{Shearing of a single discontinuity: comparison between phase-field and XFEM results.}
    \label{fig:shearing-xfem-compare}
\end{figure}

We first verify the phase-field formulation by comparing its results with those obtained by the XFEM.
To ensure that the phase-field result is accurate enough, we assign a sufficiently small length parameter with quite fine discretization, namely $L = 0.4$ mm and $L/h = 20$.
Figure~\ref{fig:shearing-xfem-compare} compares the results of the phase-field method and XFEM in terms of the shear stress and dilation in the discontinuity.
One can see that the two methods provide nearly identical results. The results show typical shear stress and dilation responses of a rough fracture undergoing shearing.
%This agreement well verifies the phase-field formulation for the single discontinuity problem.

\begin{figure}[h!]
    \centering
    \subfloat[Shear stress]{\includegraphics[height=0.4\textwidth]{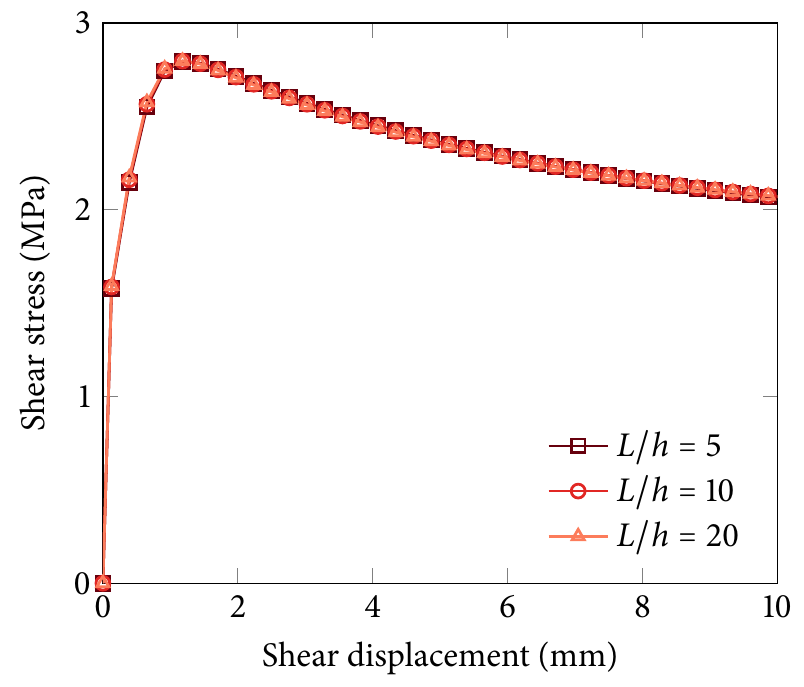}} \hspace{0.5em}
    \subfloat[Dilation]{\includegraphics[height=0.4\textwidth]{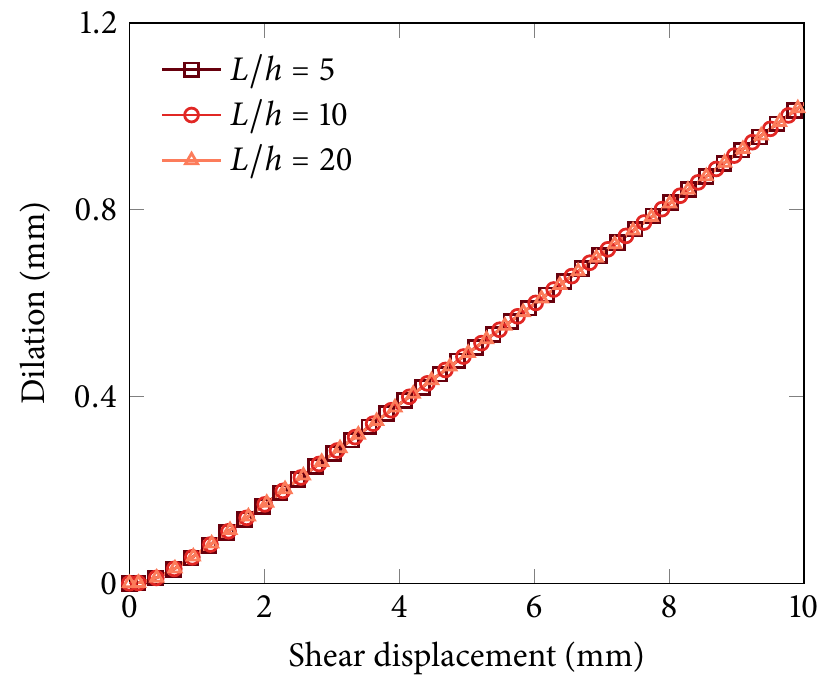}}
    \caption{Shearing of a single discontinuity: mesh sensitivity study. $L = 0.4$ mm in all cases.}
    \label{fig:shearing-mesh-convergence}
\end{figure}

Next, we examine the mesh sensitivity of the phase-field method by repeating the same problem with three different levels of refinement: $L/h=5$, $L/h=10$, and $L/h=20$.
The results are presented in Fig.~\ref{fig:shearing-mesh-convergence}.
We observe very little sensitivity to the value of $L/h$, which indicates that a rather coarse refinement level of $L/h=5$ would be good enough for practical purposes.
The results also confirm that the proposed approximation of the discontinuous strain (\cf~Eqs.~\eqref{eq:strain-f} and~\eqref{eq:strain-f-approx}), which relies on the $\Gamma$-convergence of the crack density functional, provides mesh-insensitive solutions.

\begin{figure}[h!]
    \centering
    \subfloat[Shear stress]{\includegraphics[height=0.4\textwidth]{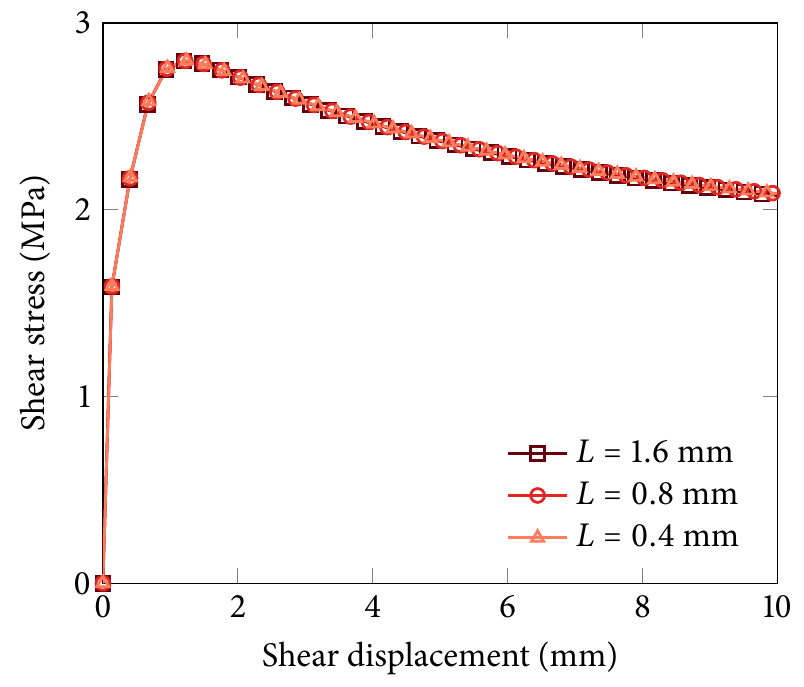}} \hspace{0.5em}
    \subfloat[Dilation]{\includegraphics[height=0.4\textwidth]{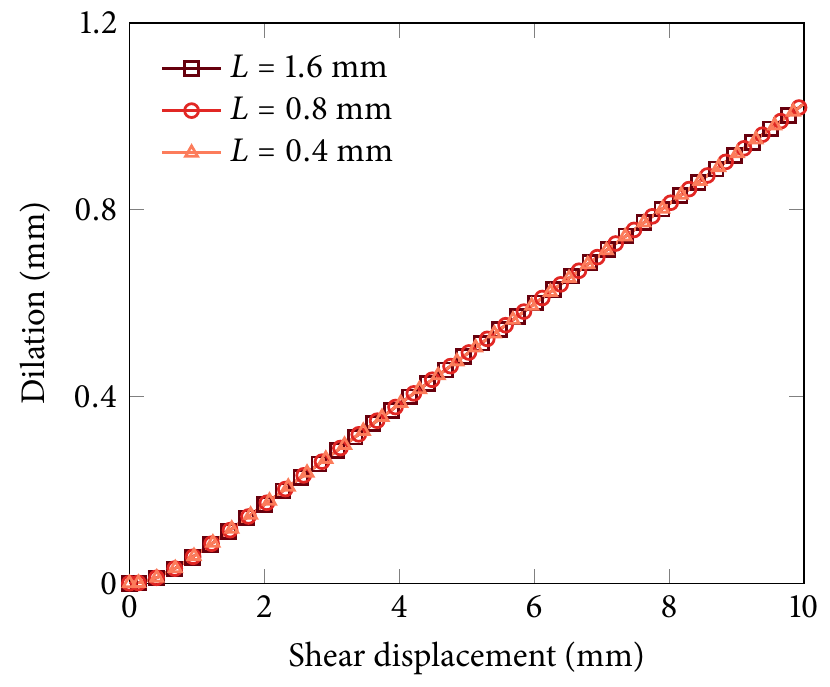}}
    \caption{Shearing of a single discontinuity: length sensitivity study. $L/h = 20$ in all cases.}
    \label{fig:shearing-L-convergence}
\end{figure}

Lastly, in Fig.~\ref{fig:shearing-L-convergence} we investigate the sensitivity of the method to the phase-field length parameter.
Here, the problem is repeated with three different length parameters, $L=1.6$ mm, $L=0.8$ mm and $L=0.4$ mm (illustrated in Fig.~\ref{fig:shearing-pf}), with a fixed refinement level of $L/h = 20$.
It can be seen that the shear stress results converge as $L$ decreases and that the dilation results are almost the same even when $L$ is fairly large at 1.6 mm.

\subsection{Biaxial compression on a rock with a single crack}
\begin{figure}[h!]
    \centering
    \includegraphics[width=0.48\textwidth]{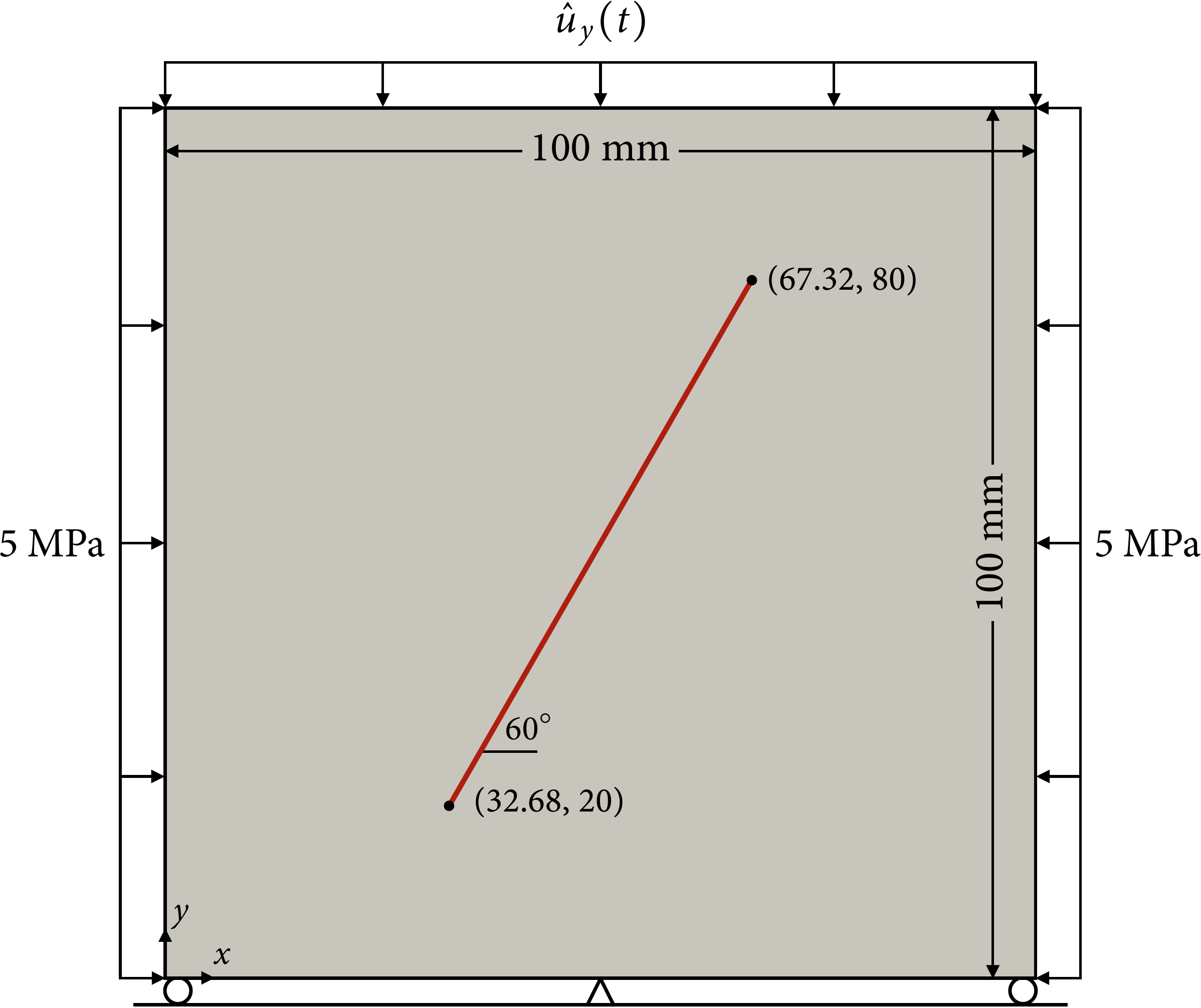}
    \caption{Biaxial compression on a rock with a single crack: geometry and boundary conditions.}
    \label{fig:biaxial-single-setup}
\end{figure}

Our second example simulates biaxial compression on a rock containing a single crack.
Figure~\ref{fig:biaxial-single-setup} illustrates the problem geometry and boundary conditions.
The domain is a 100-mm wide square, and the internal crack is inclined 60 degrees from the horizontal.
As for the boundary conditions, the domain is supported by rollers on its bottom boundary, except at the center where the displacements are constrained by a pin for stability.
To provide the crack with an initial shear strength, we apply a constant confining pressure of 5 MPa on the two lateral sides of the domain.
By default, we use the phase-field method with $L=0.2$ mm and a locally refined mesh satisfying $L/h=10$ where $d>0$.
We compress the top boundary with a prescribed uniform rate of 0.01 mm until the total compression reaches 0.1 mm after 10 steps.

\begin{figure}[h!]
  \centering
  \includegraphics[width=0.8\textwidth]{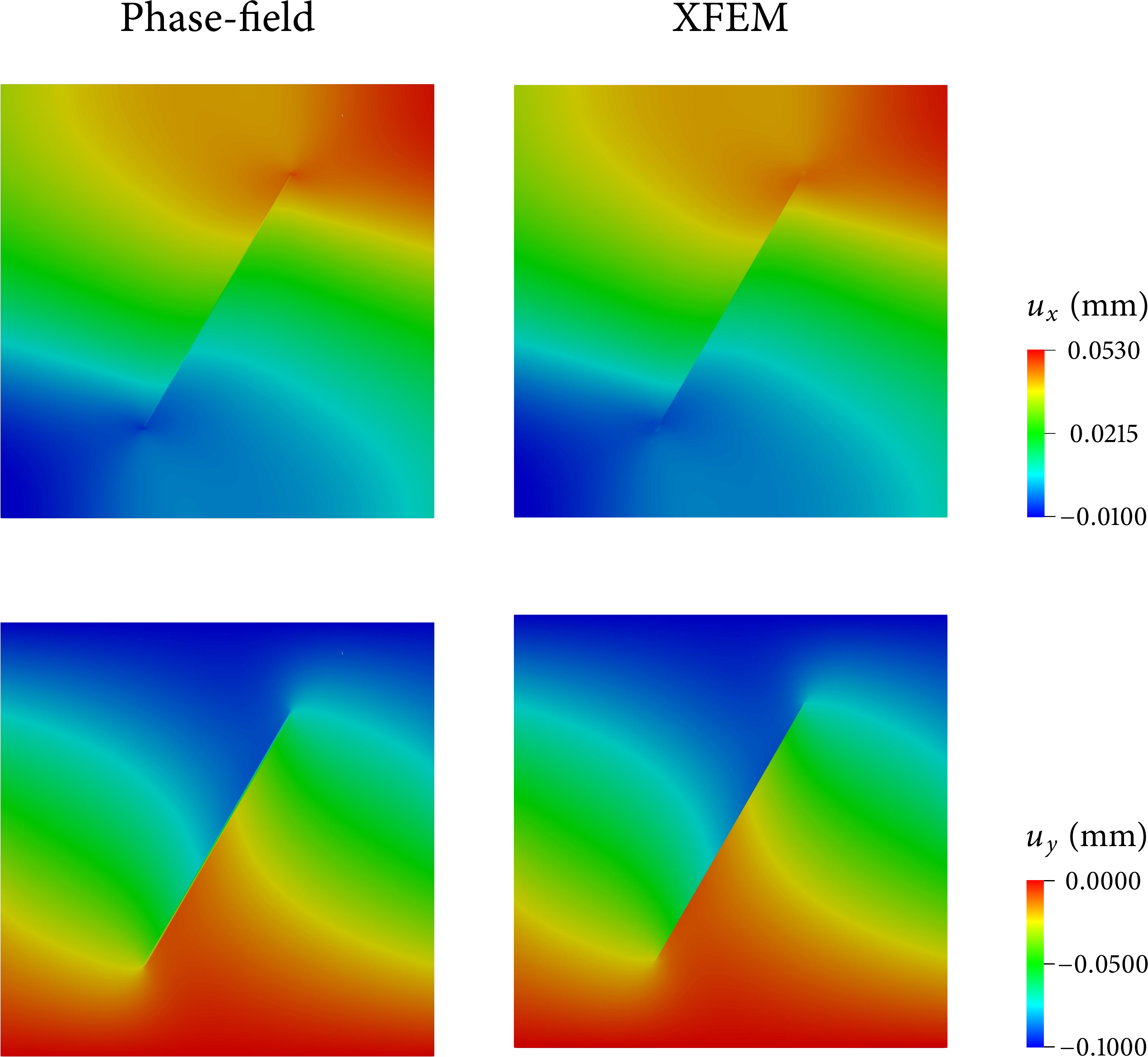}
  \caption{Biaxial compression on a rock with a single crack: comparison between phase-field and XFEM results.}
  \label{fig:single-xfem-compare}
\end{figure}

\begin{figure}[h!]
    \centering
    \subfloat[Slip]{\includegraphics[height=0.4\textwidth]{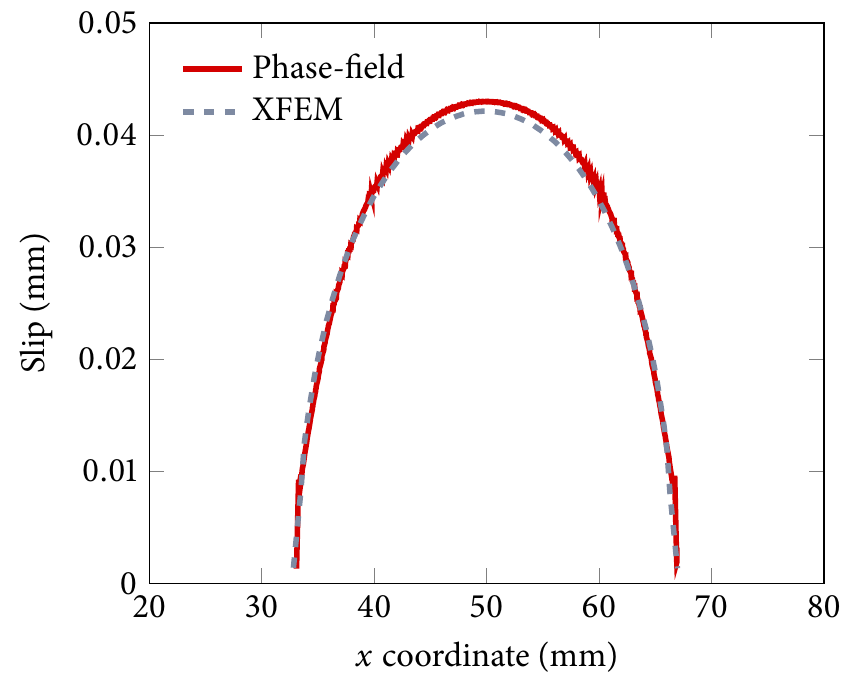}} \hspace{0.5em}
    \subfloat[Dilation]{\includegraphics[height=0.4\textwidth]{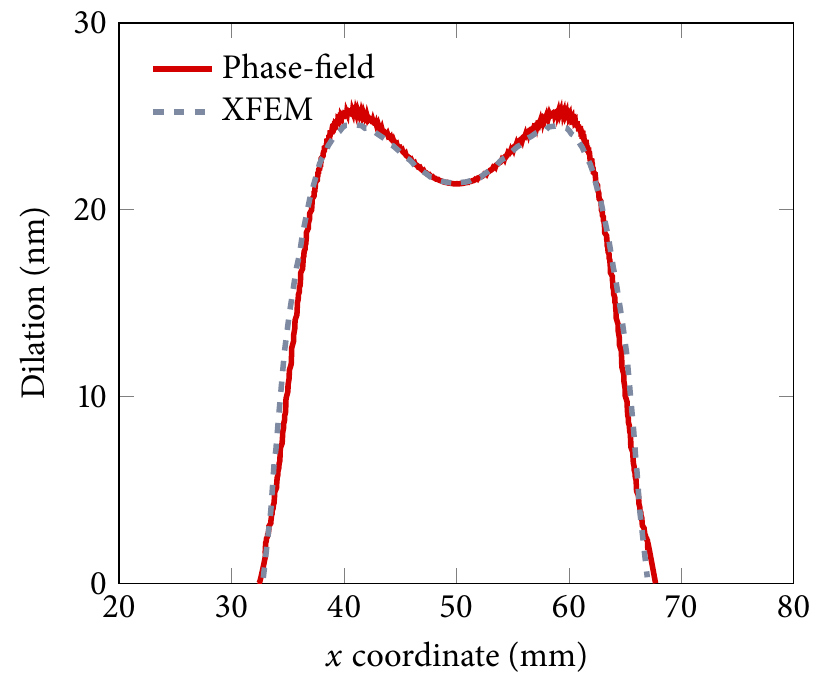}}
    \caption{Biaxial compression on a rock with a single crack: comparison between phase-field and XFEM results in the slip and dilation along the discontinuity.}
    \label{fig:single-compare-dilation-slip}
\end{figure}

Figure~\ref{fig:single-xfem-compare} compares the results obtained by the phase-field method and XFEM in terms of the $x$- and $y$-displacement fields.
The phase-field and XFEM results appear almost indistinguishable, which verifies the proposed phase-field formulation for an embedded crack problem.
For a more quantitative verification, in Fig.~\ref{fig:single-compare-dilation-slip} we further compare the two results in terms of the total slip and dilation along the crack and confirm that they match very well.
Given that the crack in this problem is not aligned with the mesh structure, these results indicate that the phase-field solution is also insensitive to the mesh alignment and thus it can be used with general unstructured meshes.

\begin{figure}[h!]
  \centering
  \subfloat[$x$-displacement]{\includegraphics[width=1.0\textwidth]{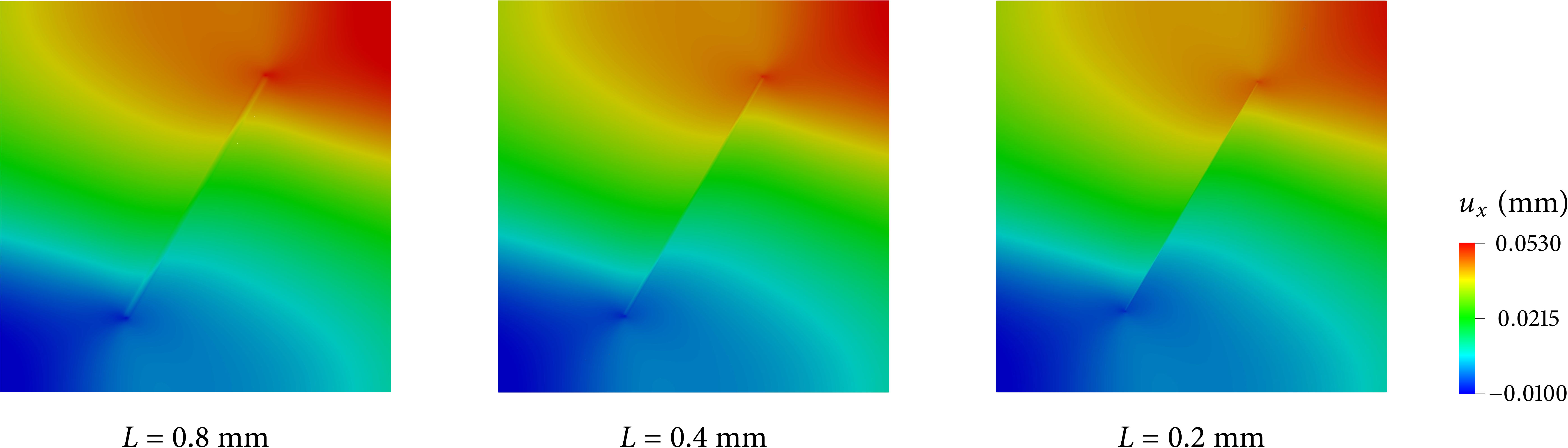}}\vspace{1em}
  \subfloat[$y$-displacement]{\includegraphics[width=1.0\textwidth]{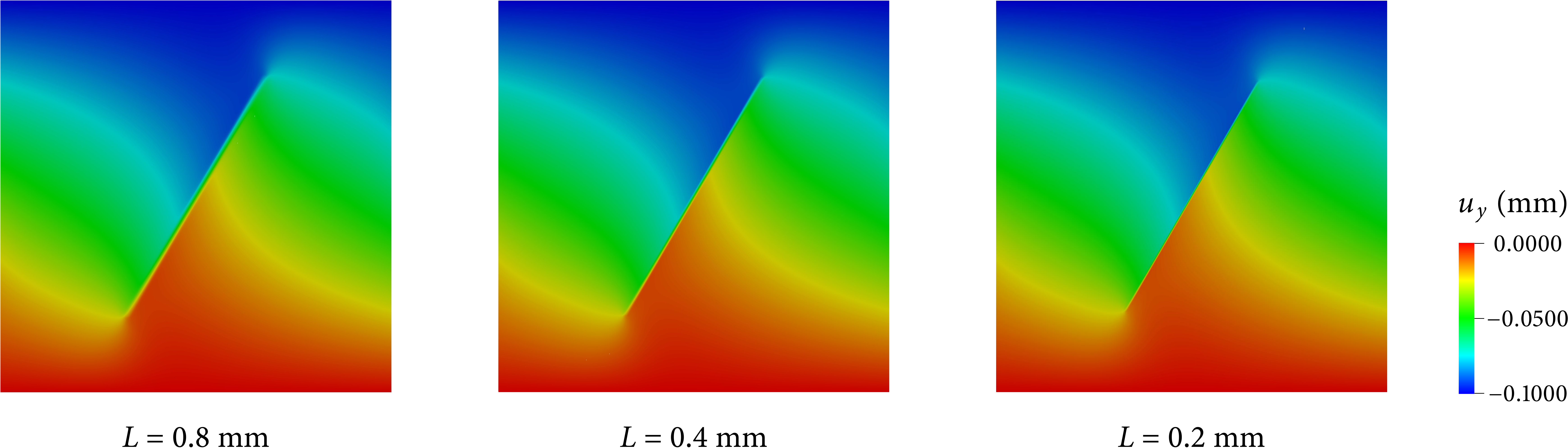}}
  \caption{Biaxial compression on a rock with a single crack: length sensitivity study.}
  \label{fig:single-L}
\end{figure}

Similar to the previous example, we examine the length sensitivity of the phase-field method by repeating this problem with three values of the length parameter: $L = 0.8$ mm, $L = 0.4$ mm and $L = 0.2$ mm.
Figure~\ref{fig:single-L} compares the $x$- and $y$-displacement fields obtained with the three length parameter values.
It can be seen that the numerical solutions are almost the same and that the displacement jump across the crack becomes sharper as the length parameter decreases.
It is thus again confirmed that the phase-field formulation is nearly insensitive to the length parameter, as long as it is reasonably small.

\subsection{Biaxial compression on a rock with two non-intersecting cracks}
\begin{figure}[h!]
    \centering
    \includegraphics[width=0.48\textwidth]{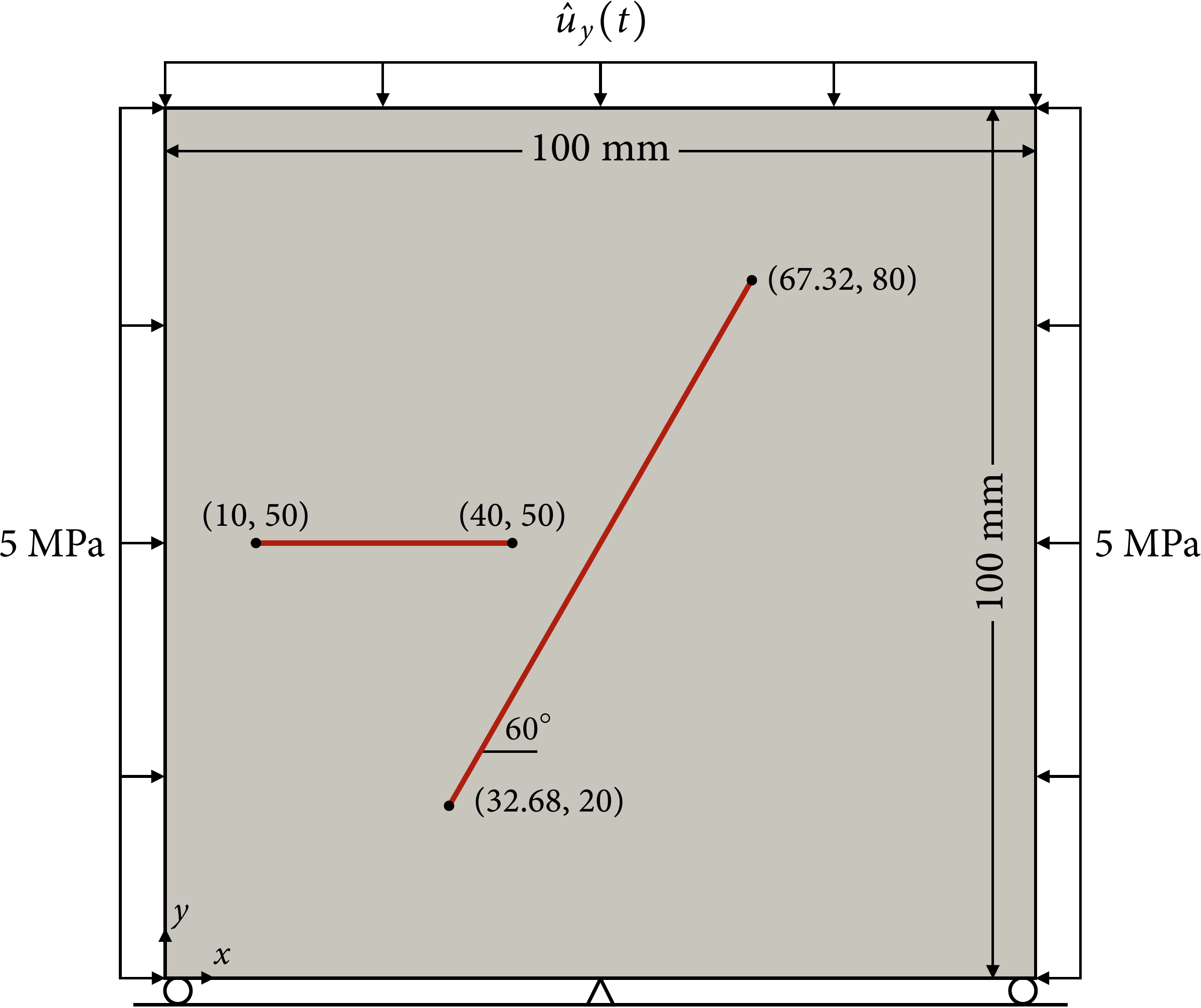}
    \caption{Biaxial compression on a rock with two non-intersecting cracks: geometry and boundary conditions.}
    \label{fig:biaxial-nonintersect-setup}
\end{figure}

We extend the previous example to a domain with two non-intersecting cracks, adding a horizontal crack to the left side of the existing one.
The specific location of the additional crack is shown in Fig.~\ref{fig:biaxial-nonintersect-setup}.
We then repeat the same biaxial compression problem, with $L = 0.2$ mm and $L/h = 10$.

\begin{figure}[h!]
  \centering
  \includegraphics[width=0.8\textwidth]{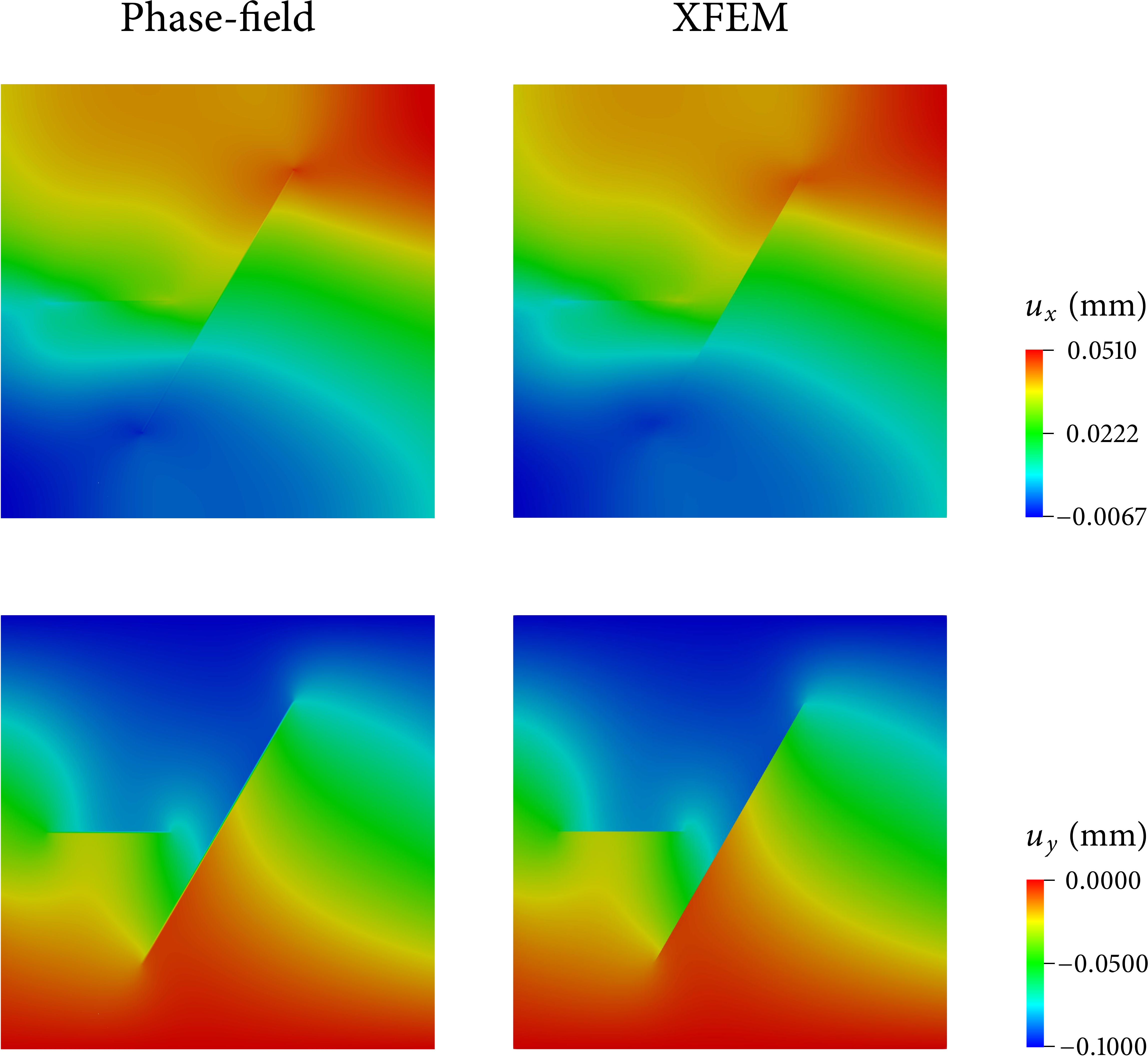}
  \caption{Biaxial compression on a rock with two non-intersecting cracks: comparison between phase-field and XFEM results.}
  \label{fig:nonintersect-xfem-compare}
\end{figure}

Figure~\ref{fig:nonintersect-xfem-compare} compares the phase-field and XFEM solutions to this problem.
We find that the phase-field and XFEM results remain nearly identical for problems with multiple discontinuities.
While not presented, we have also confirmed that the results show very little sensitivity to the phase-field length parameter as before.

\subsection{Biaxial compression on a rock with two intersecting cracks}
\begin{figure}[h!]
    \centering
    \includegraphics[width=0.48\textwidth]{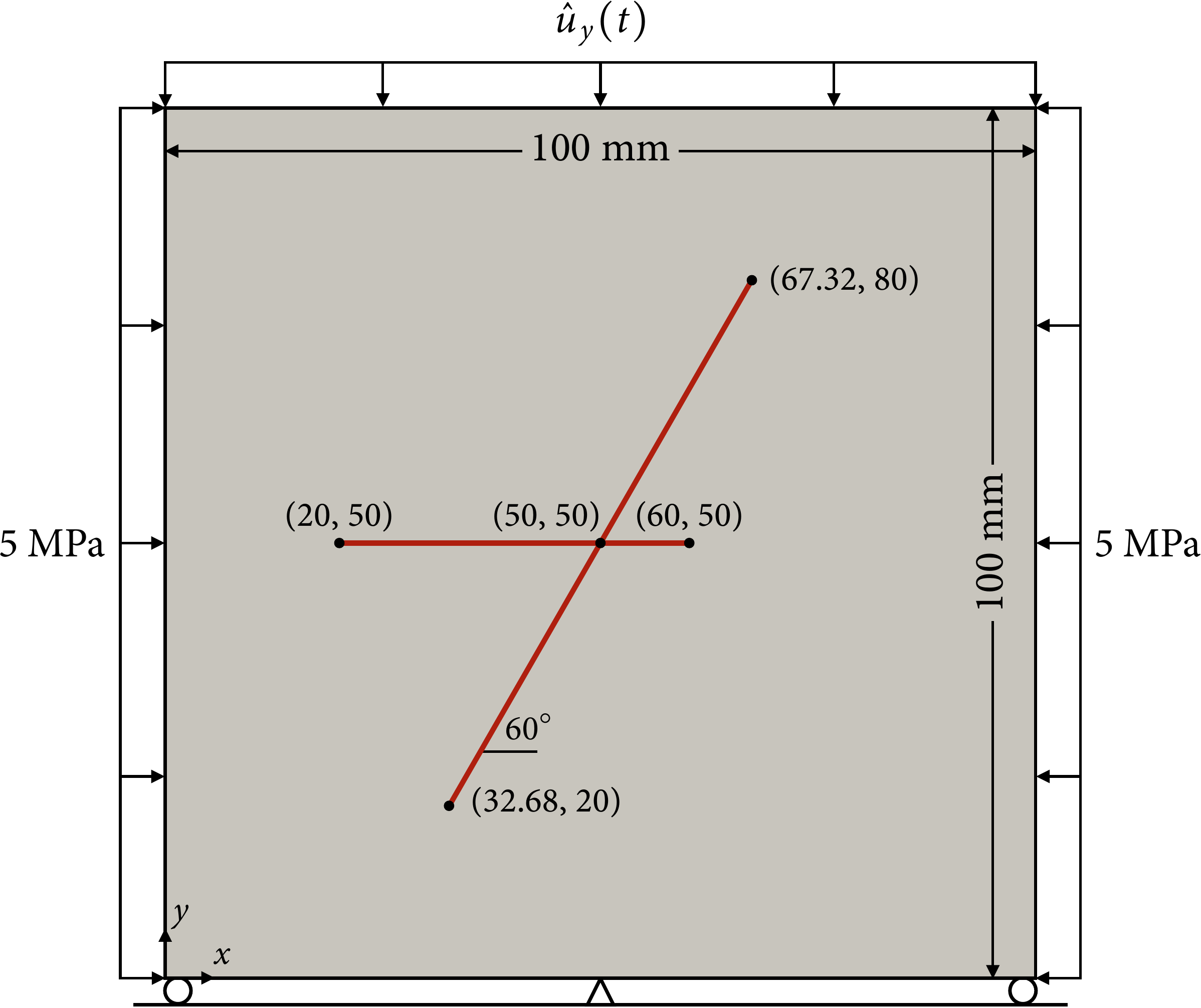}
    \caption{Biaxial compression on a rock with two intersecting cracks: geometry and boundary conditions.}
    \label{fig:biaxial-intersect-setup}
\end{figure}

As our final example for verification, we consider intersecting discontinuities---a challenging yet common scenario in geomechanics.
For this purpose, we modify the previous example by relocating and elongating the horizontal crack, as shown in Fig.~\ref{fig:biaxial-intersect-setup}.
The crack normal and tangential directions at a quadrature point near the intersection are assigned to be those of the discontinuity closer to the quadrature point at hand.

\begin{figure}[h!]
  \centering
  \includegraphics[width=0.8\textwidth]{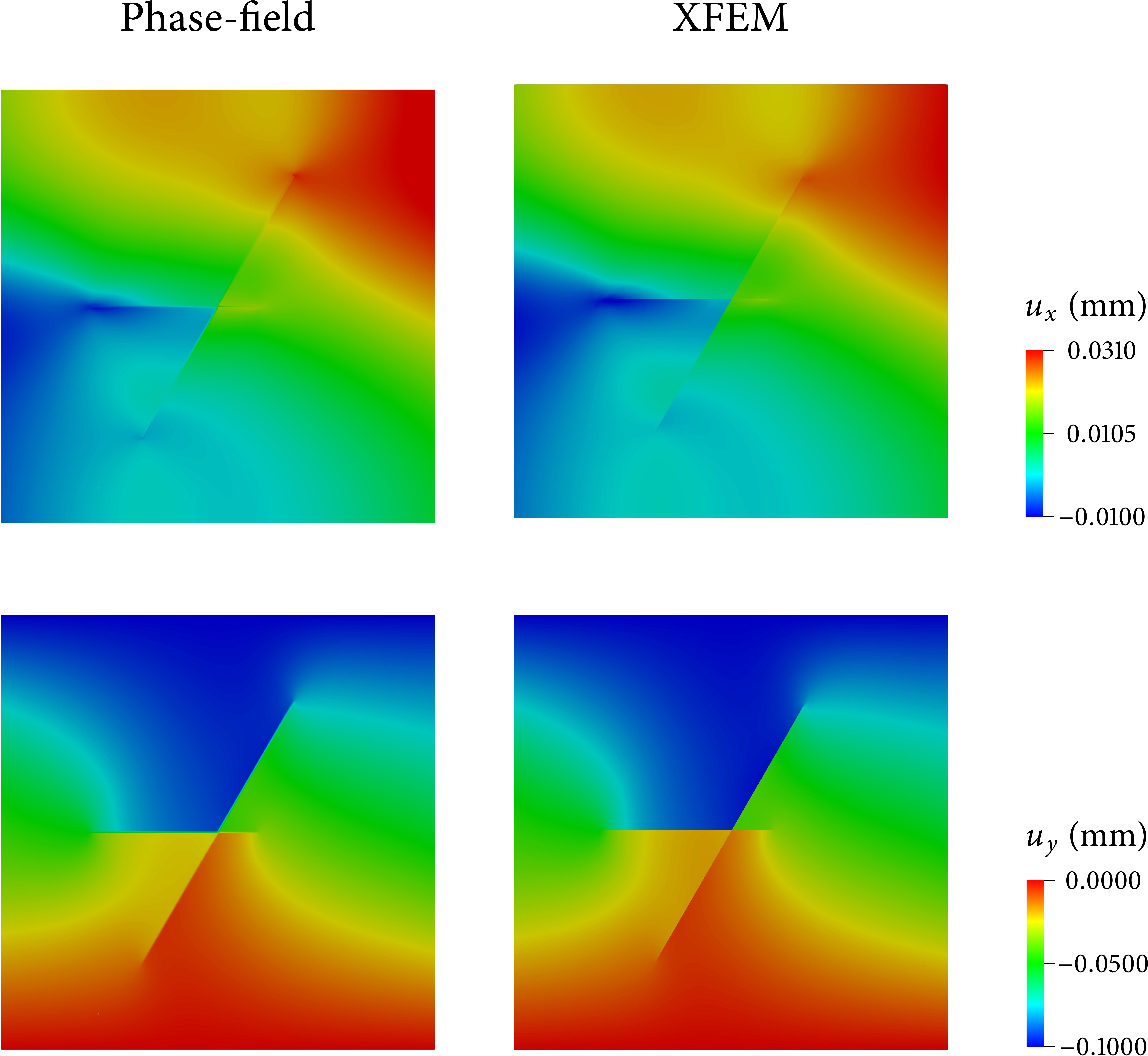}
  \caption{Biaxial compression on a rock with two intersecting cracks: comparison between the phase-field and XFEM results.}
  \label{fig:intersect-xfem-compare}
\end{figure}

The phase-field and XFEM results for this problem are compared in Fig.~\ref{fig:intersect-xfem-compare}.
The comparison shows that, even when the cracks are intersecting, the phase-field method provides a numerical solution very similar to an XFEM solution.
The simulation result is also qualitatively correct in that the intersection inhibits the slip of the inclined crack, which has also been observed by Liu et al.~\cite{liu2019modeling}.
%All of the results in this section have thus consistently verified the proposed phase-field formulation.

\begin{figure}[h!]
  \centering
  \includegraphics[width=0.5\textwidth]{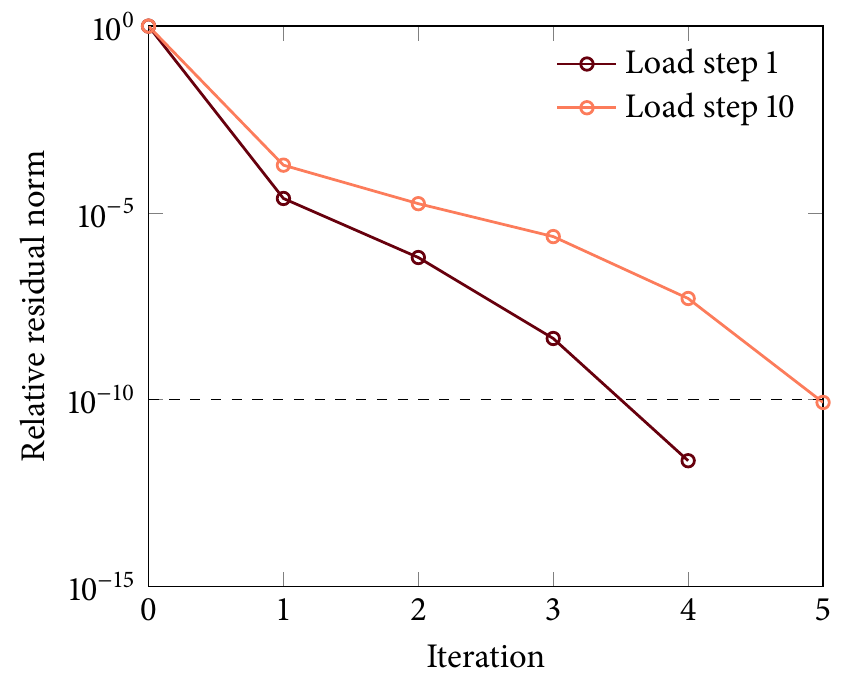}
   \caption{Biaxial compression on a rock with two intersecting cracks: Newton convergence profiles during the first load step and the tenth (last) load step.}
  \label{fig:intersect-newton}
\end{figure}

At this point, we note that Newton's method shows optimal convergence for all the numerical results in this section.
As an example, Fig.~\ref{fig:intersect-newton} shows the Newton convergence profiles during the first load step---in which the crack is in the nonlinear elastic regime---and the last load step---in which the crack is in the inelastic regime---of this problem.
Regardless of the regime, Newton's method converges after five iterations, showing asymptotically quadratic rates.
These results affirm that the tangent operators presented in Section~\ref{sec:discretization} are correct. %and that Newton's method is sufficiently robust and efficient for the phase-field formulation.

\subsection{Fracture propagation from preexisting cracks under biaxial compression}
\label{sec:fracture-propagation}

Having verified the phase-field formulation with stationary discontinuity problems, we now apply it to fracture propagation problems.
For this purpose, we extend the last two verification examples---domains with two non-intersecting cracks (Fig.~\ref{fig:biaxial-nonintersect-setup}) and two intersecting cracks (Fig.~\ref{fig:biaxial-intersect-setup})---by allowing cracks to nucleate and propagate.
We set the critical tensile fracture energy as $\mathcal{G}_{c} = 13$ J/m$^2$ and the peak tensile strength as $\stress_{p} = 3.2$ MPa.
We now solve the phase-field evolution equation~\eqref{eq:pf-evolution-residual} in each load step, with the staggered solution scheme~\cite{miehe2010phase}.
To ensure that the staggered solution is sufficiently accurate, we reduce the compression rate to $0.002$ mm. 
Other problem conditions remain unchanged. 

\begin{figure}[h!]
  \centering
  \includegraphics[width=\textwidth]{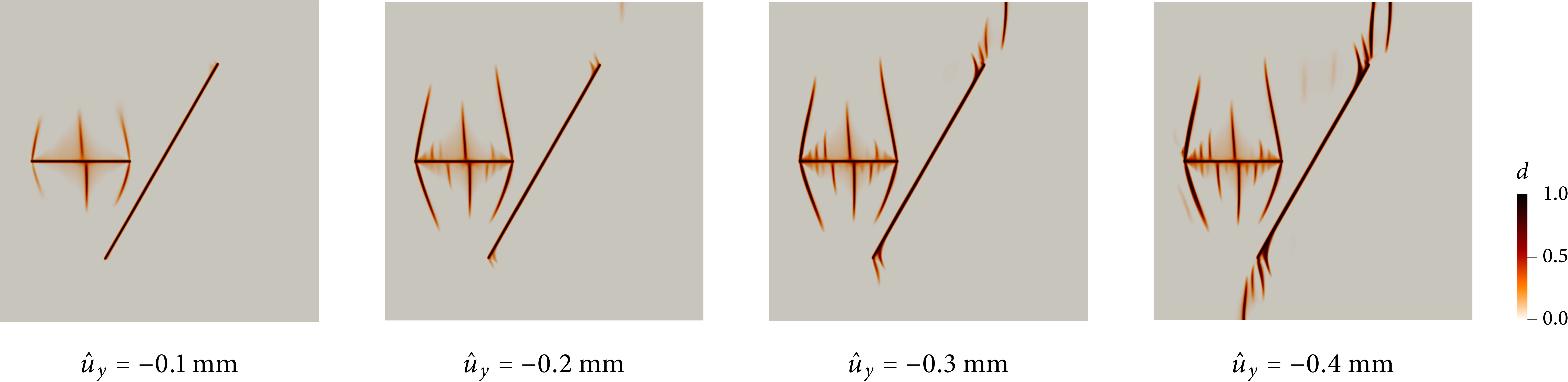}
  \caption{Fracture propagation from non-intersecting cracks under biaxial compression: phase-field evolution at different load steps.}
  \label{fig:nonintersect-cracking}
\end{figure}

\begin{figure}[h!]
  \centering
  \includegraphics[width=\textwidth]{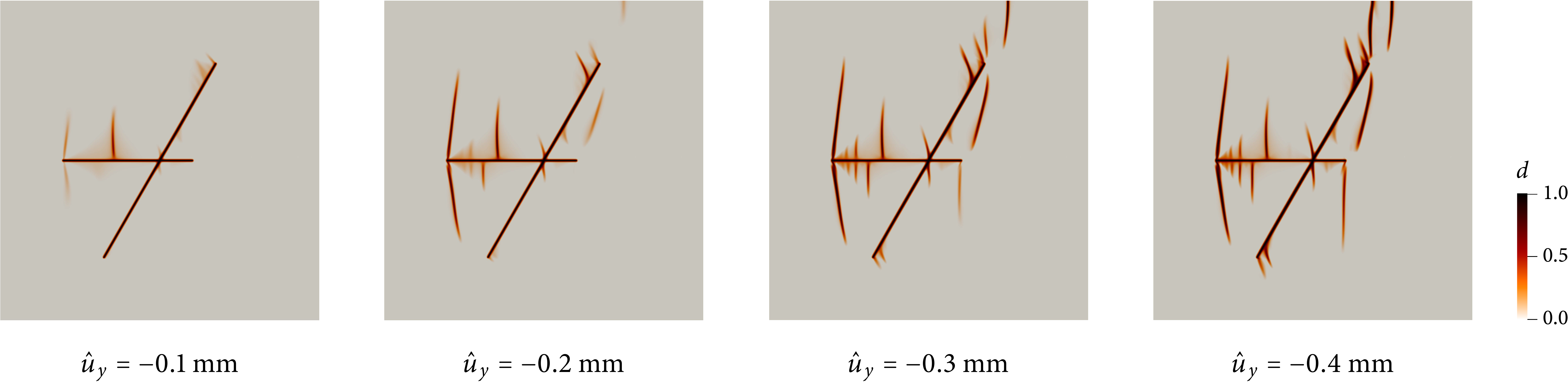}
   \caption{Fracture propagation from intersecting cracks under biaxial compression: phase-field evolution at different load steps.}
  \label{fig:intersect-cracking}
\end{figure}

Figures~\ref{fig:nonintersect-cracking} and \ref{fig:intersect-cracking} present simulation results of crack growth from non-intersecting cracks and intersecting cracks, respectively. 
In both cases, cracks emerge around the center of the horizontal crack, and wing cracks grow from the tips of the preexisting discontinuities. 
The resulting cracking patterns are highly complex, similar to experimental observations from compression tests on rock specimens with preexisting rough discontinuities (see Fig.~8 in Asadizadeh \etal~\cite{asadizadeh2019mechanical} for example). 
The ability to simulate such complex cracking patterns without any surface tracking algorithm is indeed the main advantage of the phase-field method over discrete methods. 

% SECTION 6
% ------------------------------------------------------------------------------
\section{Closure}
\label{sec:closure}

We have developed the first framework for incorporating roughness-induced deformation behavior of rock discontinuities in phase-field modeling.
The key idea is to transform a displacement-jump-based discrete constitutive model for discontinuities---the standard approach in rock mechanics---into a strain-based continuous model. 
No additional parameter is introduced in this transformation.
We then cast the strain-based constitutive model into a phase-field formulation for frictional interfaces.
It has been verified that the proposed phase-field framework provides nearly identical numerical solutions to those obtained by a well-established discrete method, for a variety of problems ranging from shearing of a single discontinuity to compression of a fractured rock with intersecting cracks.
Also demonstrated is the capabilities of the phase-field formulation for simulating complex crack growth from rough discontinuities, without any algorithm to explicitly represent crack geometry.
This work has thus constructed an unprecedented bridge between discrete constitutive models for rough discontinuities and state-of-the-art phase field methods.

% AUTHOR CONTRIBUTIONS
% ------------------------------------------------------------------------------
\section*{Author Contributions} 
\label{sec:credit}

\textbf{Fan Fei}: Conceptualization, Methodology, Software, Validation, Formal Analysis,  Writing - Original Draft, Visualization.
\textbf{Jinhyun Choo}: Conceptualization, Methodology, Software, Validation,  Writing - Original Draft, Writing - Review \& Editing, Supervision, Project Administration, Funding Acquisition.
\textbf{Chong Liu}: Software, Validation. 
\textbf{Joshua A. White}: Methodology, Software, Writing - Review \& Editing.

% ACKNOWLEDGEMENT
% ------------------------------------------------------------------------------
\section*{Acknowledgments}
This work was supported by the Research Grants Council of Hong Kong through Projects 17201419 and 27205918. 
FF received additional financial support from a Hong Kong Ph.D. Fellowship. 
JAW acknowledges financial support from TotalEnergies through the FC-MAELSTROM project. Portions of this work were performed under the auspices of the U.S. Department of Energy by Lawrence Livermore National Laboratory under Contract DE-AC52-07-NA27344.

% DATA AVAILABILITY STATEMENT
% ------------------------------------------------------------------------------
\section*{Data Availability Statement} 
\label{sec:data-availability} 

The data that support the findings of this study are available from the corresponding author upon reasonable request.

% REFERENCES
% ------------------------------------------------------------------------------
% \section*{References}
\bibliography{references}

\begin{thebibliography}{10}
\expandafter\ifx\csname url\endcsname\relax
  \def\url#1{\texttt{#1}}\fi
\expandafter\ifx\csname urlprefix\endcsname\relax\def\urlprefix{URL }\fi
\expandafter\ifx\csname href\endcsname\relax
  \def\href#1#2{#2} \def\path#1{#1}\fi

\bibitem{barton2000tbm}
N.~R. Barton, TBM tunnelling in jointed and faulted rock, CRC Press, 2000.

\bibitem{el2009deep}
S.~El~Bedoui, Y.~Guglielmi, T.~Lebourg, J.-L. P{\'e}rez, Deep-seated failure
  propagation in a fractured rock slope over 10,000 years: the la clapi{\`e}re
  slope, the south-eastern french alps, Geomorphology 105~(3-4) (2009)
  232--238.

\bibitem{borja2016rock}
R.~I. Borja, J.~Choo, J.~A. White, {Rock moisture dynamics, preferential flow,
  and the stability of hillside slopes}, in: Multi-Hazard Approaches to Civil
  Infrastructure Engineering, 2016, pp. 443--464.

\bibitem{white2014geomechanical}
J.~A. White, L.~Chiaramonte, S.~Ezzedine, W.~Foxall, Y.~Hao, A.~Ramirez,
  W.~McNab, {Geomechanical behavior of the reservoir and caprock system at the
  In Salah CO$_2$ storage project}, Proceedings of the National Academy of
  Sciences 111~(24) (2014) 8747--8752.

\bibitem{barton2020review}
N.~Barton, A review of mechanical over-closure and thermal over-closure of rock
  joints: Potential consequences for coupled modelling of nuclear waste
  disposal and geothermal energy development, Tunnelling and Underground Space
  Technology 99 (2020) 103379.

\bibitem{lepillier2020variational}
B.~Lepillier, K.~Yoshioka, F.~Parisio, R.~Bakker, D.~Bruhn, Variational
  phase-field modeling of hydraulic fracture interaction with natural fractures
  and application to enhanced geothermal systems, Journal of Geophysical
  Research: Solid Earth 125~(7) (2020) e2020JB019856.

\bibitem{fu2021close}
P.~Fu, M.~Schoenball, J.~B. Ajo-Franklin, C.~Chai, M.~Maceira, J.~P. Morris,
  H.~Wu, H.~Knox, P.~C. Schwering, M.~D. White, et~al., {Close observation of
  hydraulic fracturing at EGS Collab Experiment 1: Fracture trajectory,
  microseismic interpretations, and the role of natural fractures}, Journal of
  Geophysical Research: Solid Earth 126~(7) (2021) e2020JB020840.

\bibitem{fu2021thermo}
P.~Fu, X.~Ju, J.~Huang, R.~R. Settgast, F.~Liu, J.~P. Morris,
  Thermo-poroelastic responses of a pressure-driven fracture in a carbon
  storage reservoir and the implications for injectivity and caprock integrity,
  International Journal for Numerical and Analytical Methods in Geomechanics
  45~(6) (2021) 719--737.

\bibitem{bourdin2008variational}
B.~Bourdin, G.~A. Francfort, J.~J. Marigo, {The variational approach to
  fracture}, Journal of Elasticity 91 (2008) 5--148.

\bibitem{miehe2010thermodynamically}
C.~Miehe, F.~Welschinger, M.~Hofacker, Thermodynamically consistent phase-field
  models of fracture: Variational principles and multi-field fe
  implementations, International journal for numerical methods in engineering
  83~(10) (2010) 1273--1311.

\bibitem{borden2012phase}
M.~J. Borden, C.~V. Verhoosel, M.~A. Scott, T.~J. Hughes, C.~M. Landis, A
  phase-field description of dynamic brittle fracture, Computer Methods in
  Applied Mechanics and Engineering 217 (2012) 77--95.

\bibitem{lee2016pressure}
S.~Lee, M.~F. Wheeler, T.~Wick, Pressure and fluid-driven fracture propagation
  in porous media using an adaptive finite element phase field model, Computer
  Methods in Applied Mechanics and Engineering 305 (2016) 111--132.

\bibitem{zhang2017modification}
X.~Zhang, S.~W. Sloan, C.~Vignes, D.~Sheng, A modification of the phase-field
  model for mixed mode crack propagation in rock-like materials, Computer
  Methods in Applied Mechanics and Engineering 322 (2017) 123--136.

\bibitem{bryant2018mixed}
E.~C. Bryant, W.~Sun, A mixed-mode phase field fracture model in anisotropic
  rocks with consistent kinematics, Computer Methods in Applied Mechanics and
  Engineering 342 (2018) 561--584.

\bibitem{choo2018coupled}
J.~Choo, W.~Sun, {Coupled phase-field and plasticity modeling of geological
  materials: From brittle fracture to ductile flow}, Computer Methods in
  Applied Mechanics and Engineering 330 (2018) 1--32.

\bibitem{ha2018liquid}
S.~J. Ha, J.~Choo, T.~S. Yun, {Liquid CO$_2$ fracturing: Effect of fluid
  permeation on the breakdown pressure and cracking behavior}, Rock Mechanics
  and Rock Engineering 51~(11) (2018) 3407--3420.

\bibitem{fei2021double}
F.~Fei, J.~Choo, Double-phase-field formulation for mixed-mode fracture in
  rocks, Computer Methods in Applied Mechanics and Engineering 376 (2021)
  113655.

\bibitem{barton1977shear}
N.~Barton, V.~Choubey, The shear strength of rock joints in theory and
  practice, Rock mechanics 10~(1-2) (1977) 1--54.

\bibitem{barton1982modelling}
N.~Barton, Modelling rock joint behavior from in situ block tests: implications
  for nuclear waste repository design, Vol. 308, Office of Nuclear Waste
  Isolation, Battelle Project Management Division, 1982.

\bibitem{barton1985strength}
N.~Barton, S.~Bandis, K.~Bakhtar, Strength, deformation and conductivity
  coupling of rock joints, in: International journal of rock mechanics and
  mining sciences \& geomechanics abstracts, Vol.~22, Elsevier, 1985, pp.
  121--140.

\bibitem{fei2020phasea}
F.~Fei, J.~Choo, A phase-field method for modeling cracks with frictional
  contact, International Journal for Numerical Methods in Engineering 121~(4)
  (2020) 740--762.

\bibitem{fei2020phaseb}
F.~Fei, J.~Choo, A phase-field model of frictional shear fracture in geologic
  materials, Computer Methods in Applied Mechanics and Engineering 369 (2020)
  113265.

\bibitem{palmer1973growth}
A.~C. Palmer, J.~R. Rice, The growth of slip surfaces in the progressive
  failure of over-consolidated clay, Proceedings of the Royal Society of
  London. A. Mathematical and Physical Sciences 332~(1591) (1973) 527--548.

\bibitem{bryant2021phase}
E.~C. Bryant, W.~Sun, Phase field modeling of frictional slip with slip
  weakening/strengthening under non-isothermal conditions, Computer Methods in
  Applied Mechanics and Engineering 375 (2021) 113557.

\bibitem{olsson2001improved}
R.~Olsson, N.~Barton, An improved model for hydromechanical coupling during
  shearing of rock joints, International Journal of Rock Mechanics and Mining
  Sciences 38~(3) (2001) 317--329.

\bibitem{hakso2019relation}
A.~Hakso, M.~Zoback, The relation between stimulated shear fractures and
  production in the barnett shale: Implications for unconventional oil and gas
  reservoirs, Geophysics 84~(6) (2019) B461--B469.

\bibitem{petty2013improving}
S.~Petty, Y.~Nordin, W.~Glassley, T.~T. Cladouhos, M.~Swyer, {Improving
  geothermal project economics with multi-zone stimulation: results from the
  Newberry Volcano EGS demonstration}, in: Proceedings of the Thirty-Eighth
  Workshop on Geothermal Reservoir Engineering, 2013, pp. 11--13.

\bibitem{gischig2015hydro}
V.~S. Gischig, G.~Preisig, et~al., Hydro-fracturing versus hydro-shearing: a
  critical assessment of two distinct reservoir stimulation mechanisms, in:
  13th ISRM International Congress of Rock Mechanics, International Society for
  Rock Mechanics and Rock Engineering, 2015.

\bibitem{rinaldi2019joint}
A.~P. Rinaldi, J.~Rutqvist, {Joint opening or hydroshearing? Analyzing a
  fracture zone stimulation at Fenton Hill}, Geothermics 77 (2019) 83--98.

\bibitem{heuze1981new}
F.~E. Heuze, T.~G. Barbour, New models for rock joints and interfaces, ASTM
  Geotechnical Testing Journal 108~(GT5) (1981).

\bibitem{gens1990constitutive}
A.~Gens, I.~Carol, E.~Alonso, A constitutive model for rock joints formulation
  and numerical implementation, Computers and Geotechnics 9~(1-2) (1990) 3--20.

\bibitem{saeb1990modelling}
S.~Saeb, B.~Amadei, Modelling joint response under constant or variable normal
  stiffness boundary conditions, in: International Journal of Rock Mechanics
  and Mining Sciences \& Geomechanics Abstracts, Vol.~27, Pergamon, 1990, pp.
  213--217.

\bibitem{plesha1987constitutive}
M.~E. Plesha, Constitutive models for rock discontinuities with dilatancy and
  surface degradation, International journal for numerical and analytical
  methods in geomechanics 11~(4) (1987) 345--362.

\bibitem{white2014anisotropic}
J.~A. White, Anisotropic damage of rock joints during cyclic loading:
  constitutive framework and numerical integration, International Journal for
  Numerical and Analytical Methods in Geomechanics 38~(10) (2014) 1036--1057.

\bibitem{wu2017unified}
J.-Y. Wu, A unified phase-field theory for the mechanics of damage and
  quasi-brittle failure, Journal of the Mechanics and Physics of Solids 103
  (2017) 72--99.

\bibitem{geelen2019phase}
R.~J. Geelen, Y.~Liu, T.~Hu, M.~R. Tupek, J.~E. Dolbow, A phase-field
  formulation for dynamic cohesive fracture, Computer Methods in Applied
  Mechanics and Engineering 348 (2019) 680--711.

\bibitem{lorentz2011convergence}
E.~Lorentz, S.~Cuvilliez, K.~Kazymyrenko, Convergence of a gradient damage
  model toward a cohesive zone model, Comptes Rendus M{\'e}canique 339~(1)
  (2011) 20--26.

\bibitem{lorentz2011gradient}
E.~Lorentz, V.~Godard, {Gradient damage models: Toward full-scale
  computations}, Computer Methods in Applied Mechanics and Engineering 200~(21)
  (2011) 1927--1944.

\bibitem{gerasimov2019penalization}
T.~Gerasimov, L.~De~Lorenzis, On penalization in variational phase-field models
  of brittle fracture, Computer Methods in Applied Mechanics and Engineering
  354 (2019) 990--1026.

\bibitem{belytschko2001arbitrary}
T.~Belytschko, N.~Mo{\"e}s, S.~Usui, C.~Parimi, Arbitrary discontinuities in
  finite elements, International Journal for Numerical Methods in Engineering
  50~(4) (2001) 993--1013.

\bibitem{rivas2019two}
E.~Rivas, M.~Parchei-Esfahani, R.~Gracie, A two-dimensional extended finite
  element method model of discrete fracture networks, International Journal for
  Numerical Methods in Engineering 117~(13) (2019) 1263--1282.

\bibitem{cusini2021simulation}
M.~Cusini, J.~A. White, N.~Castelletto, R.~R. Settgast, Simulation of coupled
  multiphase flow and geomechanics in porous media with embedded discrete
  fractures, International Journal for Numerical and Analytical Methods in
  Geomechanics 45~(5) (2021) 563--584.

\bibitem{asadizadeh2019mechanical}
M.~Asadizadeh, M.~F. Hossaini, M.~Moosavi, H.~Masoumi, P.~Ranjith, Mechanical
  characterisation of jointed rock-like material with non-persistent rough
  joints subjected to uniaxial compression, Engineering Geology 260 (2019)
  105224.

\bibitem{miehe2010phase}
C.~Miehe, M.~Hofacker, F.~Welschinger, A phase field model for rate-independent
  crack propagation: Robust algorithmic implementation based on operator
  splits, Computer Methods in Applied Mechanics and Engineering 199~(45-48)
  (2010) 2765--2778.

\bibitem{steinke2019phase}
C.~Steinke, M.~Kaliske, A phase-field crack model based on directional stress
  decomposition, Computational Mechanics 63~(5) (2019) 1019--1046.

\bibitem{regueiro2001plane}
R.~A. Regueiro, R.~I. Borja, Plane strain finite element analysis of pressure
  sensitive plasticity with strong discontinuity, International Journal of
  Solids and Structures 38~(21) (2001) 3647--3672.

\bibitem{verhoosel2013phase}
C.~V. Verhoosel, R.~de~Borst, A phase-field model for cohesive fracture,
  International Journal for Numerical Methods in Engineering 96~(1) (2013)
  43--62.

\bibitem{miehe2015minimization}
C.~Miehe, S.~Mauthe, S.~Teichtmeister, Minimization principles for the coupled
  problem of darcy--biot-type fluid transport in porous media linked to phase
  field modeling of fracture, Journal of the Mechanics and Physics of Solids 82
  (2015) 186--217.

\bibitem{mauthe2017hydraulic}
S.~Mauthe, C.~Miehe, Hydraulic fracture in poro-hydro-elastic media, Mechanics
  Research Communications 80 (2017) 69--83.

\bibitem{choo2018cracking}
J.~Choo, W.~Sun, Cracking and damage from crystallization in pores: Coupled
  chemo-hydro-mechanics and phase-field modeling, Computer Methods in Applied
  Mechanics and Engineering 335 (2018) 347--379.

\bibitem{bandis1983fundamentals}
S.~Bandis, A.~Lumsden, N.~Barton, Fundamentals of rock joint deformation, in:
  International Journal of Rock Mechanics and Mining Sciences \& Geomechanics
  Abstracts, Vol.~20, Pergamon, 1983, pp. 249--268.

\bibitem{simo1990class}
J.~C. Simo, M.~Rifai, A class of mixed assumed strain methods and the method of
  incompatible modes, International journal for numerical methods in
  engineering 29~(8) (1990) 1595--1638.

\bibitem{son2004elasto}
B.-K. Son, Y.-K. Lee, C.-I. Lee, Elasto-plastic simulation of a direct shear
  test on rough rock joints, International Journal of Rock Mechanics and Mining
  Sciences 41 (2004) 354--359.

\bibitem{nagel2016advantages}
T.~Nagel, U.-J. G{\"o}rke, K.~Moerman, O.~Kolditz, {On advantages of the Kelvin
  mapping in finite element implementations of deformation processes},
  Environmental Earth Sciences 75~(11) (2016) 937.

\bibitem{de2011computational}
E.~A. de~Souza~Neto, D.~Peric, D.~R. Owen, Computational Methods for
  Plasticity: Theory and Applications, John Wiley \& Sons, 2011.

\bibitem{cuitino1992material}
A.~Cuitino, M.~Ortiz, A material-independent method for extending stress update
  algorithms from small-strain plasticity to finite plasticity with
  multiplicative kinematics, Engineering computations (1992).

\bibitem{de1994note}
R.~de~Borst, A.~E. Groen, {A note on the calculation of consistent tangent
  operators for von Mises and Drucker-Prager plasticity}, Communications in
  numerical methods in engineering 10~(12) (1994) 1021--1025.

\bibitem{moes1999finite}
N.~Mo{\"e}s, J.~Dolbow, T.~Belytschko, A finite element method for crack growth
  without remeshing, International journal for numerical methods in engineering
  46~(1) (1999) 131--150.

\bibitem{liu2008contact}
F.~Liu, R.~I. Borja, A contact algorithm for frictional crack propagation with
  the extended finite element method, International Journal for Numerical
  methods in engineering 76~(10) (2008) 1489--1512.

\bibitem{wibowo1994effect}
J.~Wibowo, B.~Amadei, S.~Sture, R.~Price, Effect of boundary conditions on the
  strength and deformability of replicas of natural fractures in welded tuff:
  Data analysis, Tech. rep., Sandia National Laboratories, Albuquerque, NM
  (United States); University of Colorado Boulder, CO (United States). (1994).

\bibitem{arndt2021deal}
D.~Arndt, W.~Bangerth, D.~Davydov, T.~Heister, L.~Heltai, M.~Kronbichler,
  M.~Maier, J.-P. Pelteret, B.~Turcksin, D.~Wells, {The deal. II finite element
  library: Design, features, and insights}, Computers \& Mathematics with
  Applications 81 (2021) 407--422.

\bibitem{choo2016hydromechanical}
J.~Choo, J.~A. White, R.~I. Borja, {Hydromechanical modeling of unsaturated
  flow in double porosity media}, International Journal of Geomechanics 16~(6)
  (2016) D4016002.

\bibitem{white2016block}
J.~A. White, N.~Castelletto, H.~A. Tchelepi, {Block-partitioned solvers for
  coupled poromechanics: A unified framework}, Computer Methods in Applied
  Mechanics and Engineering 303 (2016) 55--74.

\bibitem{choo2018large}
J.~Choo, Large deformation poromechanics with local mass conservation: An
  enriched {G}alerkin finite element framework, International Journal for
  Numerical Methods in Engineering 116 (2018) 66--90.

\bibitem{choo2019stabilized}
J.~Choo, {Stabilized mixed continuous/enriched Galerkin formulations for
  locally mass conservative poromechanics}, Computer Methods in Applied
  Mechanics and Engineering 357 (2019) 112568.

\bibitem{lee2001influence}
H.~Lee, Y.~Park, T.~Cho, K.~You, Influence of asperity degradation on the
  mechanical behavior of rough rock joints under cyclic shear loading,
  International Journal of Rock Mechanics and Mining Sciences 38~(7) (2001)
  967--980.

\bibitem{annavarapu2014nitsche}
C.~Annavarapu, M.~Hautefeuille, J.~E. Dolbow, {A Nitsche stabilized finite
  element method for frictional sliding on embedded interfaces. Part I: single
  interface}, Computer Methods in Applied Mechanics and Engineering 268 (2014)
  417--436.

\bibitem{choo2021barrier}
J.~Choo, Y.~Zhao, Y.~Jiang, M.~Li, C.~Jiang, K.~Soga, A barrier method for
  frictional contact on embedded interfaces, arXiv preprint arXiv:2107.05814
  (2021).

\bibitem{liu2019modeling}
C.~Liu, J.~H. Pr{\'e}vost, N.~Sukumar, Modeling branched and intersecting
  faults in reservoir-geomechanics models with the extended finite element
  method, International Journal for Numerical and Analytical Methods in
  Geomechanics 43~(12) (2019) 2075--2089.

\end{thebibliography}
% \printbibliography

% APPENDIX
% ------------------------------------------------------------------------------
\appendix

\section{Derivatives in return mapping}
\label{appendix:derivates}
This appendix describes how to calculate the derivatives in the return mapping algorithm described in Section~\ref{sec:return-mapping}.
Firstly, the derivatives of the yield and potential functions (Eqs.~\eqref{eq:yield-function} and \eqref{eq:potential-function}, respectively) are given by
\begin{linenomath}
\begin{align}
  \dfrac{\pd F}{\pd \tstress_{f}} & = \dfrac{1}{2} \tensor{\alpha} + \stress_{\cn,n} \dfrac{\pd \tan (\phi_\mathrm{b} + \omega \psi )}{\pd \tstress_{f}}  , \label{eq:grad-F} \\
  \dfrac{\pd^2 G}{\pd \tstress_{f} \dyadic \pd \tstress_{f}} &=  \dfrac{\pd \tan \psi}{\pd \tstress_{f}}\dyadic (\tensor{n} \dyadic \tensor{n}) . \label{eq:hess-G} \\
  \dfrac{\pd^{2} G}{\pd \tstress_{f} \pd \Delta \lambda} &= \dfrac{\pd \tan \psi}{\pd \Delta \lambda} (\tensor{n} \dyadic \tensor{n}) .
\end{align}
\end{linenomath}
In the specific constitutive model employed herein~\cite{white2014anisotropic}, the friction angle and dilation angle are independent of stress. Thus,
\begin{linenomath}
\begin{align}
  \dfrac{\pd \tan(\phi_\mathrm{b} + \omega \psi)}{\pd \tstress_{f}} &= \tensor{0} , \\
  \dfrac{\pd \tan \psi}{\pd \tstress_{f}}  &= \tensor{0} .
\end{align}
\end{linenomath}
In this case, Eqs~\eqref{eq:grad-F} and \eqref{eq:hess-G} simplify to
\begin{linenomath}
\begin{align}
  \dfrac{\pd F}{\pd \tstress_{f}} & = \dfrac{1}{2} \tensor{\alpha}  , \label{eq:grad-F-white} \\
  \dfrac{\pd^2 G}{\pd \tstress_{f} \dyadic \pd \tstress_{f}} &=  \tensor{0} . \label{eq:hess-G-white}
\end{align}
\end{linenomath}
Next, we calculate the derivatives of the friction and dilation angles with respect to the discrete plastic multiplier $\Delta \lambda$.
Using chain rule, we get
\begin{linenomath}
\begin{align}
  \dfrac{\pd \tan (\phi_\mathrm{b} + \omega \psi)}{\pd \Delta \lambda} &= \dfrac{\pd \tan (\phi_\mathrm{b} + \omega \psi)}{\pd v^{\pl}_{f}} \dfrac{\pd v^{\pl}_{f}}{\pd \lambda} \dfrac{\pd \lambda}{\pd \Delta \lambda} , \label{eq:friction-delta-lambda} \\
  \dfrac{\pd \tan  \psi}{\pd \Delta \lambda} &= \dfrac{\pd \tan \psi}{\pd v^{\pl}_{f}} \dfrac{\pd v^{\pl}_{f}}{\pd \lambda} \dfrac{\pd \lambda}{\pd \Delta \lambda} . \label{eq:dilation-delta-lambda}
\end{align}
\end{linenomath}
Considering the discrete form of the plastic multiplier, \ie~$\lambda = \lambda_{n} + \Delta \lambda$, we can simplify the above equations as
\begin{linenomath}
\begin{align}
  \dfrac{\pd \tan (\phi_\mathrm{b} + \omega \psi)}{\pd \Delta \lambda} &= \dfrac{\pd \tan (\phi_\mathrm{b} + \omega \psi)}{\pd v^{\pl}_{f}} \dfrac{\pd v^{\pl}_{f}}{\pd \lambda} , \label{eq:friction-lambda} \\
  \dfrac{\pd \tan  \psi}{\pd \Delta \lambda} &= \dfrac{\pd \tan \psi}{\pd v^{\pl}_{f}} \dfrac{\pd v^{\pl}_{f}}{\pd \lambda} . \label{eq:dilation-lambda}
\end{align}
\end{linenomath}
Combining the flow rule in Eq.~\eqref{eq:flow-rule} and Eq.~\eqref{eq:v-p} gives
\begin{linenomath}
\begin{align}
  \dfrac{\pd v^{\pl}_{f}}{\pd \lambda} = \dfrac{1}{\Gamma_d(d, \grad d)} \, .
\end{align}
\end{linenomath}
Inserting the above equation into Eqs.~\eqref{eq:friction-lambda} and \eqref{eq:dilation-lambda} gives
\begin{linenomath}
\begin{align}
  \dfrac{\pd \tan (\phi_\mathrm{b} + \omega \psi)}{\pd \Delta \lambda} &= \dfrac{1}{\Gamma_d(d, \grad d)} \dfrac{\pd \tan (\phi_\mathrm{b} + \omega \psi)}{\pd v^{\pl}_{f}} , \\
  \dfrac{\pd \tan  \psi}{\pd \Delta \lambda} &= \dfrac{1}{\Gamma_d(d, \grad d)} \dfrac{\pd \tan  \psi}{\pd v^{\pl}_{f}} .
\end{align}
\end{linenomath}
For the particular constitutive model we employed~\cite{white2014anisotropic}, the specific expressions of the above equations are given by (\cf~Eqs.~\eqref{eq:white-dilation}--\eqref{eq:white-Lambda})
\begin{linenomath}
\begin{align}
	\dfrac{\pd \tan (\phi_\mathrm{b} + \omega \psi)}{\pd v^{\pl}_{f}} &= \dfrac{1}{\cos^{2} (\phi_\mathrm{b} + \omega \psi)} \left(\omega \dfrac{\pd \psi}{\pd v^{\pl}_{f}} + \psi \dfrac{\pd \omega}{\pd v^{\pl}_{f}} \right)  , \\
	\dfrac{\pd \tan  \psi}{\pd v^{\pl}_{f}} &=  \dfrac{\tan \psi_\mathrm{r}}{b}\dfrac{1}{\cosh^2 (v^{\pl}_{f}/ b)} ,
\end{align}
\end{linenomath}
where
\begin{linenomath}
\begin{align}
	\dfrac{\pd \psi}{\pd v^{\pl}_{f}} &= \dfrac{1}{1 + \tan^2 \psi} \dfrac{\pd \tan \psi}{\pd v^{\pl}_{f}}	, \\
	\dfrac{\pd \omega}{\pd v^{\pl}_{f}} &= (1 - \omega_{0})c\exp(-c \Lambda) .
\end{align}
\end{linenomath}

\end{document}